\documentclass{aa}  
\usepackage{graphicx}
\usepackage{natbib}
\usepackage{txfonts}
\usepackage{caption}
\usepackage{xcolor}
\usepackage{soul}
\begin{document}
   \title{Analysis of apsidal motion in eclipsing binaries using TESS data}

   \subtitle{II. A test of internal stellar structure}

\author{    A. Claret~\inst{1,2}
       \and  A. Gim\'enez  \inst{3,4}
       \and D. Baroch \inst {5,6} 
       \and I. Ribas  \inst {5,6}
       \and J. C. Morales  \inst {5,6}
       \and G. Anglada-Escudé \inst{5,6} }

   \offprints{A. Claret, e-mail:claret@iaa.es}
\institute{Instituto de Astrof\'{\i}sica de Andaluc\'{\i}a, CSIC, Apartado 3004,
 18080 Granada, Spain
 \and
 Dept. F\'{\i}sica Te\'{o}rica y del Cosmos, Universidad de Granada, 
 Campus de Fuentenueva s/n,  10871, Granada, Spain\
    \and Centro de Astrobiologia (CSIC-INTA), E-28850 Torrej\'on de Ardoz, Madrid, Spain 
     \and International Space Science Institute (ISSI), Hallerstrasse 6, CH-3012 Bern, Switzerland
     \and  Institut de Ci\`encies de l’Espai (ICE, CSIC), Campus UAB, c/ Can Magrans s/n, 
     E-08193 Bellaterra, Barcelona, Spain
     \and
      Institut d’Estudis Espacials de Catalunya (IEEC), c/ Gran Capit\`a 2-4, E-08034 Barcelona, Spain}

   \date{Received; accepted; }

 \abstract
   {The measurement of apsidal motion rates in eccentric eclipsing binaries is a unique way to gain insight 
    into the internal structure of stars through the internal density concentration parameter, $k_2$. 
    High-quality physical parameters of the stellar components, together with precise measurements of the advance 
    of the periastron, are needed for the comparison with values derived from models.}
   {As a product of the Transiting Exoplanet Survey Satellite (TESS) mission, high-precision light curves of a large number of eclipsing binaries 
    are now available. Using a selection of well-studied, double-lined eccentric eclipsing binary systems, we aim to 
    determine their apsidal motion rates and place constraints on the internal density concentration and compare with 
    the predictions from state-of-the-art theoretical models.}
    {We computed times of minimum light using the TESS light curves of 34 eclipsing binaries with precise 
    absolute parameters. We used the changing difference over time between primary and secondary eclipse timings 
    to determine the apsidal motion rate. To extend the time baseline, we combined the high-precision TESS timings 
    with reliable archival data. On the other hand, for each component of our sample of double-lined 
    eclipsing binaries, we computed grids of evolutionary stellar models for the observed stellar mass exploring  
    ranges of values of the overshooting parameter $f_{\rm ov}$, the mixing-length parameter, and the metallicity. 
    To find the best solution for the two components we adopted a $\chi^2$  statistic to infer the optimal values 
    of the overshooting parameter and the mixing-length parameter. The theoretical internal structure constants to be 
    compared with the observed values were calculated by integrating the differential equations of Radau for each 
    stellar model.}
   {We have determined the apsidal motion rate of 27 double-lined eclipsing binaries with precise 
    physical parameters. The obtained values, corrected for their relativistic contribution, yield precise 
    empirical parameters of the internal stellar density concentration. The comparison 
    of these results with the predictions based on new theoretical models shows very good agreement. 
    Small deviations are identified but remain within the observational uncertainties and the path for a 
    refined comparison is indicated. }
   {}

   \titlerunning {Apsidal Motion Test }
   \maketitle
%

\section{Introduction}

Double-lined eclipsing binaries (DLEBs) have demonstrated to be the basic source of information about fundamental stellar properties, 
such as masses and radii \citep{Andersen1991}. The comparison of observed physical properties with theoretical models has 
been used to perform critical tests of stellar structure and evolution models \citep{Ribas2000,Torres2010,Claret2019}. Precise masses 
and radii are needed for a solid and reliable comparison between observations and theory, 
essentially better than 3\%, and this generally calls for the use of DLEBs. Furthermore, the binary systems have to be well 
detached, for instance, with the radii of both components being much smaller than their Roche limits, to ensure that the components 
represent the behavior of single stars with the same physical properties.

Eccentric eclipsing binaries offer further opportunities to characterize the internal structure of stars through the measurement 
of the precession rate of the line of the apses of the orbit, for instance, the apsidal motion rate. Such secular motion can be 
understood as the sum of two terms, one classical and one relativistic. The classical, or Newtonian, effect is related to the
quadrupole interactions that depend on the internal mass distribution of the stellar components (basically the degree of mass concentration 
toward the center). The second term is a contribution arising from general relativity (GR), the best known example of which is the 
advance of the   perihelion of Mercury. Of course, if there is a third body gravitationally bound to the close binary, an additional term in the 
apsidal motion has to be considered. 

Precise determinations of apsidal motion require long-term monitoring of the times of eclipse, generally spanning several decades, with 
high-quality measurements, although apsidal motion rates can also be derived from long time series of radial  velocity data  \citep{Schmitt2016}.  The Transiting Exoplanet Survey Satellite (TESS) mission to study exoplanets through 
photometric transits \citep{Ricker2015}, with its nearly full sky coverage, provides, as a bonus, precise photometry of a large sample 
of eclipsing binary systems with a time baseline of at least 27 days  and up to two years in some cases. Precise monitoring of binary light 
curves is thus possible from space, without the disturbing day/night effect, and accurate eclipse timings can be derived thanks to the 
uniform sampling. Equipped with this new tool, we have established a program to make precise apsidal motion determinations in eccentric 
eclipsing binaries with accurate absolute dimensions, in some cases for the first time, and compared them with theoretical models.

The first results of our program were presented in \cite{Baroch2021}, hereafter referred to as paper I. This paper addressed eclipsing 
binary systems with accurate dimensions and with apsidal motion rates dominated by the relativistic contribution, with a limit set to be 
at least 60\% of the total apsidal motion rate. This allowed us to perform a stringent test of the predictions of GR, which 
revealed excellent agreement between observations and theory. In the present paper, we focus on the systems where the classical term is 
dominant (for instance, GR contribution being less than 60\% of the total). For such cases, we calculate the GR apsidal motion rate analytically 
(which paper I shows to be accurate) and we subsequently subtract it from the observed rate to determine the observational classical term. 
We can then compare with stellar model predictions and provide constraints on interior structure 
\citep[for instance,][and references therein]{Claret1993,Claret2010b}.

This paper is structured as follows. Section\,\ref{sec:sample} is dedicated to describe the observational sample and the 
measurements of eclipse timings. Section\,\ref{sec:determination} describes the apsidal motion determinations, the methodology and 
the results, with further details for each individual system in an appendix. Section\,\ref{sec:models} is dedicated to describe the 
stellar evolutionary models, the differential equations used to obtain theoretical values of the apsidal motion constants, and the 
methodology employed to compare with the physical dimensions of the component stars. Section\,\ref{sec:structure} is devoted to the 
comparison between observed and theoretical values of log k$_2$ and, finally, in Section\,\ref{sec:conclusions} we present our 
conclusions.

 \section{The observational sample} \label{sec:sample}

For a useful interpretation of the apsidal motion rate observed in eccentric eclipsing binaries, it is essential to 
have a precise knowledge of the physical properties of the component stars, essentially masses and radii. Some of the equations, 
for example, have a strong dependence on the relative radii as they contribute to the fifth power. For this reason, we have limited our  
dynamical study using eclipse timings to those cases where the masses and radii of the components are known with an accuracy better
than 3\%. A list of well-detached eccentric eclipsing binaries with good 
absolute dimensions was published by \cite{Torres2010}, and we have further added a number of systems from the DEBCAT catalog of
\cite{Southworth2015}, which is updated permanently. We have only considered systems with TESS measurements of both primary 
and secondary eclipses, thus permitting the determination of the timing difference. Due to the expected amplitude of the apsidal motion 
variations, we also set a lower limit to the orbital eccentricity at 0.01.

Our analysis is restricted to systems with an expected relativistic contribution below
60\% of the total apsidal motion. Those with a larger relativistic contribution were discussed in paper I and are less useful 
for the study of internal structure due to the larger relative uncertainty of the observed classical term. The systems analyzed 
in the present paper are listed in Table\,\ref{tab:sample}, together with their main physical parameters and the corresponding 
references, sorted by decreasing mass of the primary component. In addition to orbital period, masses and radii of the component 
stars, we also provide the effective temperatures and the projected rotational velocities, necessary for the computation 
of the apsidal motion rates, as described in section 4.

  The systems V380 Cyg (B1.1 III and B2.5/3 V), V636 Cen (G0 V and G0 V)  and CM Dra (M4.5 V and M4.5 V) are not considered due to the difficulties found with standard theoretical stellar models in reproducing their observed physical parameters, namely their masses, radii and effective temperatures, independently of the apsidal motion results. This is a requisite of our methodology to obtain theoretical apsidal motion parameters, as described in Section 4. Each of these systems has some characteristics that push it beyond the boundaries of our studied parameter space: the evolved stage and the proximity to the Roche limit of the primary component of V380 Cyg, the strong chromospheric activity of V636 Cen, or the very low masses and magnetic activity of CM Dra. Investigating these systems will require detailed individual studies of the observational data and model input physics, which is left for subsequent publications.  On the other hand, 
we have considered three systems that should have been included in paper I due to an expected relativistic contribution above 60\%, 
but did not have either sufficient TESS data at the time of publication or a reliable apsidal motion determination. These systems are
V1022 Cas, EW Ori, and BF Dra, and they are included in Table\,\ref{tab:sample} and discussed in Section \ref{sec:determination}.

For all systems in Table\,\ref{tab:sample} we have analyzed the available TESS photometric information retrieved from Sectors 1 to 34. To compute the time of minimum light of the eclipses, we first normalized the TESS light curves using the out-of-eclipse phases. We selected well-sampled individual eclipses using the same orbital phase interval for all primary and secondary eclipses, and computed their time of minimum light employing the widely-used \cite{Kwee1956} method. We then computed the corresponding difference between primary and secondary eclipse timings, $T_2-T_1$, expressed in days, 
 and listed the resulting values in Table\,\ref{tab:t2t1}, which is available electronically. The Table gives the values of $T_2-T_1$ determined from the individual timings, together with their separation in orbital cycles 
({\it dN}), that we have limited to $\pm$1.

\section{Determination of the apsidal motion rates} \label{sec:determination}

In Paper I we performed the determination of the apsidal motion rate from the analysis of the time-variation of the difference 
between primary and secondary eclipses, $T_2-T_1$. This method assumes independent knowledge of the orbital eccentricity and that the 
variations in the timing differences can be represented by a linear relationship with the slope corresponding to the time derivative of 
the argument of periastron. This method is only valid when considering a small fraction, typically less than 1\%, of the total apsidal 
motion period, $U$. Otherwise, the nonlinear component of the $T_2-T_1$ variations becomes relevant and the analysis requires a different 
approach. When this is the case, one can use the equations given by \cite{Gimenez1995}, which are complete up to $\mathcal{O}(e^5)$.
Linearizing the variations permits computing the argument of periastron, $\omega$, corresponding to each observed value of $T_2-T_1$ with 
the adopted orbital eccentricity and inclination, as derived from the light curve analysis. This is based on the relation between the phase 
of the secondary eclipse and the value of $e\cos\omega$. Potential ambiguities in the argument of periastron can be resolved with the value 
of $e\sin\omega$ resulting from the light curve analysis. Using this approach, a linear fit to the variation of $\omega$ with time yields 
the determination of the apsidal motion rate with no restriction in terms of the coverage of the period, $U$.

In both approaches, the basic observational information is the measurement of $T_2-T_1$. We provide in Table \ref{tab:t2t1} the full list of 
the $T_2-T_1$ TESS values for all systems in Table\,\ref{tab:sample}, and we discuss each individual system in Appendix A. 
In Table\,\ref{tab:apsidal} we summarize the results of the apsidal motion determination where the 
orbital eccentricity, $e$, and the apsidal motion rate, $\dot\omega$ expressed in deg/cycle, are provided. The anomalistic period, $P_a$, can 
be easily computed from the sidereal period in Table\,\ref{tab:sample}, using the relation $P_a = P_s/(1-\dot\omega_{\rm obs}/360)$. The letter code in the 
second column denotes the adopted methodology: A for those based on the linear variation of $T_2-T_1$, and B for those using the argument of 
periastron, including the comparison of values derived from the light curve analysis. In addition, we use L to denote values adopted from the 
literature, generally with method B, or using variable $\omega$ in the analysis of light curves spanning a long time base. In this case, we 
checked that the adopted solutions predict values for the relative position of the secondary eclipse in agreement with the new TESS 
measurements. Obviously, B and L determinations correspond to faster apsidal motion rates and A to slower ones, including those with a 
higher fractional contribution of the relativistic term. Finally, those systems where the presence of a third body has been proposed, either 
from the light curve analysis or from the variations of the individual eclipse timings, are marked with an asterisk.  

Table\,\ref{tab:apsidal} does not include all systems in Table\,\ref{tab:sample} because some were discarded due to various reasons. Firstly, let us note that we defined a 
precision criterion to accept a system for further analysis. We only kept systems with a relative precision better than 25\% in the classical 
term of the apsidal motion determination, for which we derive the internal stellar structure constants. This corresponds to a 
maximum $\sigma_{\log k_2}$ of 0.11, necessary for a constraining comparison with theoretical models. After our analysis, we found a small 
fraction of the eclipsing binaries in Table\,\ref{tab:sample} to yield no significant apsidal motion detections. For two systems, BP Vul and AI Phe, we were 
not able to measure apsidal motion in spite of the precise TESS observations. We encountered a similar situation with MU Cas and V1022 Cas, 
which resulted in apsidal motion rate determinations having uncertainties above our acceptance threshold because of the narrow time span covered 
by precise timings. In the case of BF Dra, we found disagreement between the different methodologies to determine the apsidal motion and could thus 
be affected by large systematic effects. Moreover, the potential presence of a perturbing third body as well as the highly-evolved nature of 
the component stars, close to the termination-age main-sequence, call for an individual study of this system, including the analysis of a new 
light curve. Finally, EW Ori does not meet the relative precision limit in the classical term because of the high relative GR contribution 
($\sim$80\%). This system could have been included in Paper I but we did not have an apsidal motion rate determination at the time. However, 
we briefly discuss the comparison with GR and the updated post-Newtonian parameters including this eclipsing binary in Appendix A.

\section{The theoretical stellar models} \label{sec:models}

We use
the Modules for Experiments in Stellar Astrophysics package \citep[MESA;][]{Paxton2011,Paxton2013,Paxton2015} version r-7385, and 
adopted the solar-calibrated value of the mixing length parameter to be 1.84 \citep{Torres2015b}. 
The  equation relating the temperature gradients was solved using the Henyey option, and for the localization of the boundary of the convection 
zone we adopted the Schwarzschild criterion. For the opacities we use the mixture of \cite{Asplund2009}, for which $Z_{\odot} = 0.0134$. 
These opacities were paired with a linear enrichment law given by $Y_p=0.249$ \citep{Ade2016} and a slope $\Delta Y/ \Delta Z = 1.67$, 
where $Y_p$ is primordial helium mass fraction. Microscopic diffusion was considered (see details below) for all evolutionary tracks 
that were computed starting from the pre-main sequence (PMS) stage. For each component of our sample of DLEBs we computed 
grids for the observed stellar mass exploring ranges of values of the overshooting parameter $f_{\rm ov}$ (see below), 
the mixing-length parameter $\alpha_{\rm MLT}$, and the metallicity $Z$. To find the solution providing the best fit to both stellar
components we adopted a $\chi^2$ statistic and considered $f_{\rm ov}$ and $\alpha_{\rm MLT}$ as the optimization parameters. We also adopted the 
initial metal abundance $Z$ to be the same for the two components. 

We calculate evolutionary tracks for the exact component masses and other parameters. Thus, our methodology does not involve 
computing grids for a wide variety of parameters and avoids interpolation in masses, metallicities, 
mixing-length, or core overshooting, which could lead to systematic effects. In our calculations we allowed the optimized ages for both 
components to differ by up to 5\%, as long as the radii, effective temperatures and 
masses were predicted to be within their respective observational error bars. This flexibility is justified by the expected uncertainties in 
the input physics of the evolutionary tracks, for instance, opacities, equations of state, mass loss, etc., and also in the observational parameters. 
In some cases this procedure was improved with some additional computations. Starting from previously converged models, we increased 
the resolution in the input physics ($\alpha_{\rm MLT}$, $f_{\rm ov}$ and $Z$) to refine the fitting process. 

A fraction of the systems in our sample have high-mass components and therefore core overshooting can have a significant effect in the 
computation of evolutionary model grids. Convective core overshooting is related to the increase of the stellar core beyond the boundary 
defined by the Schwarzschild criterion. Stellar models computed taking into account this extra-mixing lead to longer main-sequence 
lifetimes and a higher degree of mass concentration toward the center, with direct impact on the comparison between theoretical and 
observational rates of apsidal motion. \citet[][and references therein]{Claret2019} selected 50 well-measured detached DLEBs to calibrate 
the dependency of $f_{\rm ov}$ on stellar mass. In such a formulation, the extra-mixture is modeled as a diffusive process 
\citep{Freytag1996,Herwig1997} with a diffusion coefficient at a radial distance $r$ from the boundary  given by 
$D(r) = D_{\rm o} \exp (-2r/f_{\rm ov} H_p)$, where $D_{\rm o}$ is the coefficient inside the boundary and $H_p$ the 
pressure scale height. As a result of the comparison between theoretical models and observational data 
from these systems, it has been shown that $f_{\rm ov}$ increases sharply up to a mass on the order of 2.0 M$_{\odot}$ 
and is practically constant up to 4.43 M$_{\odot}$, the upper limit in mass of the observational sample   adopted by \cite{Claret2019}. 
There are also some evidences of a dependence of core overshooting on stellar mass from a theoretical point of view \citep[see, for 
example, the appendices in][]{Claret2017,Jermyn2018}. Such kind of calibrations can only be performed with highly evolved 
stars where the effects of core overshooting are more evident. Unfortunately, this is not the case in our observational sample with 
only two systems with moderately evolved components (V453 Cyg and V478 Cyg). For all stars more massive than 10 M$_{\odot}$, we have 
extended the range of values of $f_{\rm ov}$ to be explored with our search methodology.

In all our computations we have included microscopic diffusion (8 elements: H$^1$, He$^3$, He$^4$, C$^{12}$, N$^{14}$, O$^{16}$, Ne$^{20}$, 
and Mg$^{24}$). For those stars showing convective envelopes, we adopted the standard mixing-length formalism \citep{Vitense1958}. For the 
most massive stars in our sample, we have adopted the \cite{Wink2001} formulation for mass loss assuming a multiplicative scale factor 
$\eta = 0.1$ (not to be confused with $\eta$ of the Radau equation; see below). On the other hand, rotational-mixing was not included in 
the calculations. For a more complete description of our modeling framework, see \cite{Claret2019} and references therein.

We also ran a consistency check of our model comparisons with observational data (masses, T$_{\rm eff}$, $\log g$, $Z$) between our adopted
MESA models and those computed with the GRANADA code \citep{Claret2004}. This was done for a subsample of systems (DI Her, V1143 Cyg, 
V1647 Sgr, IQ Per, $\zeta$ Phe). Convective core overshooting in the GRANADA code is treated as a step-function, characterized by 
the parameter $\alpha_{\rm ov}$. A similar check-up procedure was already carried out by \cite{Claret2016,Claret2017} for a larger 
sample of DLEBs. From this cross-comparison the relationship $\alpha_{\rm ov}/f_{\rm ov} \approx 11.36 \pm0.22$ was derived.  
With this procedure we check the consistency between the log k$_2$ computed with MESA and with the GRANADA codes. A very good agreement 
has been found between the k$_2$ computed with the two codes.

To calculate the theoretical values of log k$_2$ to be compared with the corresponding empirically-determined values, we applied 
the methodology described above to all systems in Table\,\ref{tab:sample}.
The theoretical internal structure constants were computed for each component through the integration of the differential equations 
of Radau of order $j$, namely
\begin{equation}
 {a {d\eta_{j}(a)\over da}}+ {6\rho(a)\over\overline\rho(a)}{(\eta_{j}\!+\!1)}+
 {\eta_{j}(\eta_{j}\!-\!1)} = {j(j+1}), \, j=2,3,4, 
\end{equation}
\noindent
where
\begin{equation}
\eta \equiv {{a}\over{\epsilon_{j}}} {d\epsilon_{j}\over{da}}
\end{equation}
\noindent
and $a$ is the mean radius of the equipotential, $\epsilon_j$ is a measure of the deviation from sphericity (the tesseral harmonics),  
$\rho(a)$ is the mass density at the distance {\it a} from the center, and $\overline\rho(a)$ is the mean mass density within an 
equipotential of radius {\it a}. The boundary conditions are: $\eta_{j}(0)$ =$j-2$, and 
$\left(d\eta_{j}\over{da}\right)_{\eta=0}$ = $-{3(j-1)\over{j+1}}{dD\over{da}}$, where $D={\rho/{\overline\rho}}$.

The theoretical internal structure constant k$_j$, for each component $j$, is given by 
\begin{equation}
 k_{j} = {{j +1 - \eta_{j}(R)}\over{2\left(j+\eta_{j}(R)\right)}},
\end{equation}
\noindent
where $\eta_{j}(R)$ are the values of $\eta_{j}$ at the surface of the star. Finally, the theoretical value of k$_2$ to be 
compared with observations is given by the weighted average over the model predictions for the two components:
\begin{equation}
\label{eq:k2_mean}
{\overline k_{2, {\rm theo}}} = {{c_{21}{k_{21, {\rm theo}}}+{c_{22}{k_{22, {\rm theo}}}}}\over {c_{2, 1} + c_{2, 2}}}. 
\end{equation}
The parameters c$_{2i}$, for the case of aligned rotation axes, are computed following the equations,
\begin{equation}
\label{eq:C_weights}
{c_{2 i}} \!=\! \left[\! \left({\Omega_{i}\over{\Omega_{K}}}\right)^{\!\!2}
\!\!\left(1 + {m_{3-i}\over{m_{i}}}\right)\!f(e)  +
{15\, m_{3-i}\over{m_{i}}}\,g(e)
\right]\!\!
{\left({R_{i}\over{A}}\right)^{\!\!5}}
\end{equation}
\noindent
where $A$ is the semi-major axis of the orbit, while $f$ and $g$ are auxiliary functions of the orbital eccentricity given by 
\begin{equation}
{f(e)} = {(1 - e^2)^{-2}}
\end{equation}
\noindent
and 
\begin{equation}
{g(e)} = {(8 + 12e^2 + e^4) f(e)^{2.5}\over{8}}.
\end{equation}
\noindent
Additionally, ${\Omega_{i}/{\Omega_{K}}}$  is the ratio between the rotational angular velocity of component $i$ and 
the average orbital angular velocity, $m_i$ are the  stellar masses, $R_i$ are the stellar radii and the eccentricity is denoted by the 
symbol $e$. As indicated by \cite{Claret1993}, the contribution of the terms $k_n$ (for $n>2$) are negligible in the present context.

On the other hand, using the quasi-spherical approximation, \cite{Claret1999} found that the influence of rotation on internal structure 
depends on the distortion of the configuration following the expression $\delta \log k_2$ $\approx -0.87 \Lambda_s + 0.004$, where 
$\Lambda_s =  2v^2/(3gR)$, with $v$ being the rotational velocity, $g$ the local gravity and $R$ the stellar radius. Such correction was 
computed at the surface of each star in our sample, using the observed rotational velocities given in Table\,\ref{tab:sample}, and applied to the 
corresponding theoretical models. We note that this correction does not take into account the effects of rotational-mixing on the mass 
concentration. It is also important to point out some uncertainty in relation to the observed rotational velocities regarding whether or not 
they are representative of the stellar interior. Most of the systems in our observational sample are not very evolved, from which it follows 
that the angular velocity gradients in their interiors should not be high. This circumstance reduces the uncertainties although it does 
not remove them completely.  

The effects of stellar compressibility and dynamic tides were also taken into account according to the computations by \cite{Claret2002}. 
We could verify that in all systems in their Table 2, no significant effect is present except in the most massive and close systems, 
EM Car and Y Cyg, for which we have applied the calculated correction (2\% and 1\%, respectively).

\begin{figure}
 \includegraphics[height=8.cm,width=6cm,angle=-90]{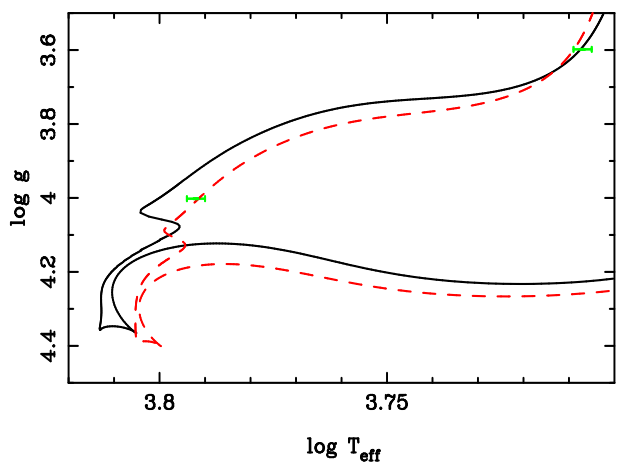}
 \caption{HR diagram for AI Phe. The models were calculated adopting $Z = 0.011$. The solid line indicates the primary component while 
the dashed one represents the secondary.}
\label{fig:AIPhe_HR}
\end{figure}

\begin{figure}
 \includegraphics[height=8.cm,width=6cm,angle=-90]{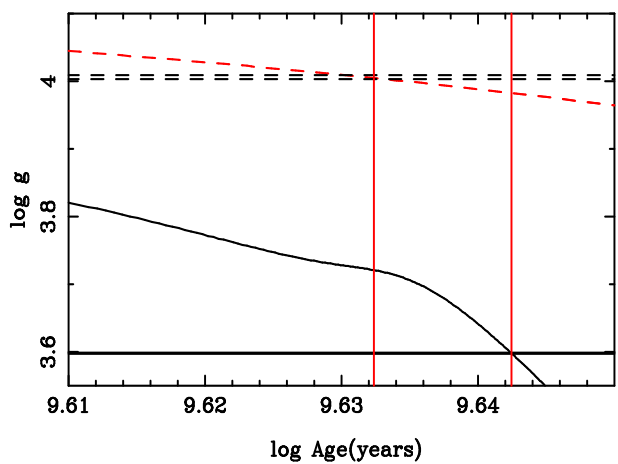}
 \caption{$\log g$ and respective error bars as a function of time for AI Phe. The two vertical lines indicate the time interval corresponding 
to the individual ages for each component. The horizontal lines represent the error bars in log g.  Line coding as in Fig.\,\ref{fig:AIPhe_HR}. }
\label{fig:AIPhe-age}
\end{figure}

\begin{figure}
 \includegraphics[height=8.cm,width=6cm,angle=-90]{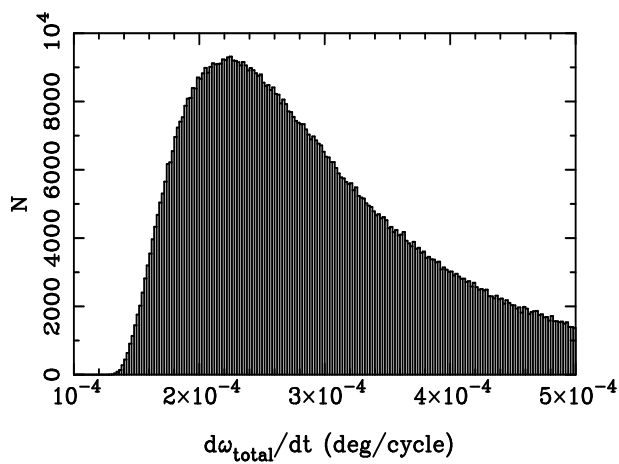}
 \caption{Monte Carlo simulations ($10^6$ realizations) for the theoretical apsidal motion of AI Phe. }
 \label{fig:AIPhe_MC}
\end{figure}

As an example of our procedure, we present in Fig.\,\ref{fig:AIPhe_HR} the fitting process in the case of AI Phe. The best input physics for the corresponding models and for each component (subindices 1 and 2) is given by $\alpha_{\rm MLT, 1} = 2.00$, $\alpha_{\rm MLT, 2} = 2.10$ and 
$f_{\rm ov1} = f_{\rm ov2} = 0$. In Fig.\,\ref{fig:AIPhe-age} we show the
resulting models in the $\log g$ vs. $\log {\rm age}$ diagram. The observational error bars are very small and were shown in different color for the sake of clarity. 
This solution, as explained above, is obtained on the basis of the minimum $\chi^2$ of the model grid search and imply the log k$_{2 i}$ values for the system. 
In the case of AI Phe, we find $\log k_{2, 1} = -1.58\pm0.30$ and $\log k_{2, 2} = -2.45\pm0.03$. 
Although the masses of both components are similar, the large difference between apsidal motion constants is due to the advanced 
evolutionary stage of the system. While the secondary is still in the main sequence, the primary is much more evolved and is located in 
the giant branch, where $k_2$ varies very rapidly, and hence the large error bar. As we do not yet have a reliable determination of the 
apsidal motion rate for AI Phe, we have performed Monte Carlo simulations  (see section 5.1 for more details), assuming that the two components are aligned with the orbital spin 
(see Fig.\,\ref{fig:AIPhe_MC}). The asymmetry of the simulations reflect the differences between the uncertainty in $\log k_2$ for each of the 
components. The most probable value of $\dot\omega_{\rm theo} \approx 2.24_{-0.50}^{+1.12}\times10^{-4}$ degrees/cycle, should be compared with 
the observational value when available, but the slow predicted rate is consistent with the lack of a measurable value
(see Appendix A). We have selected AI Phe to illustrate our methodology, in spite of not having an observed value 
of the apsidal motion rate, as a tribute to the late Prof. J. Andersen. This was one of his favorite DLEBs, frequently used as a test case 
for stellar structure and evolution.

\section {The apsidal motion test of stellar structure} \label{sec:structure}

The weighted values of the model-predicted $\log k_2$ given by Eq.\,(\ref{eq:k2_mean}) are presented in column 6 of Table\,\ref{tab:apsidal}, together with the corresponding 
predicted apsidal motion rates. The adopted methodology ensures that both components have the same age (within the previously mentioned 5\% tolerance limit), and that the observed masses, radii and effective temperatures, are reproduced by the respective models within their errors.

\begin{figure}
 \includegraphics[height=8.cm,width=16cm,angle=-90]{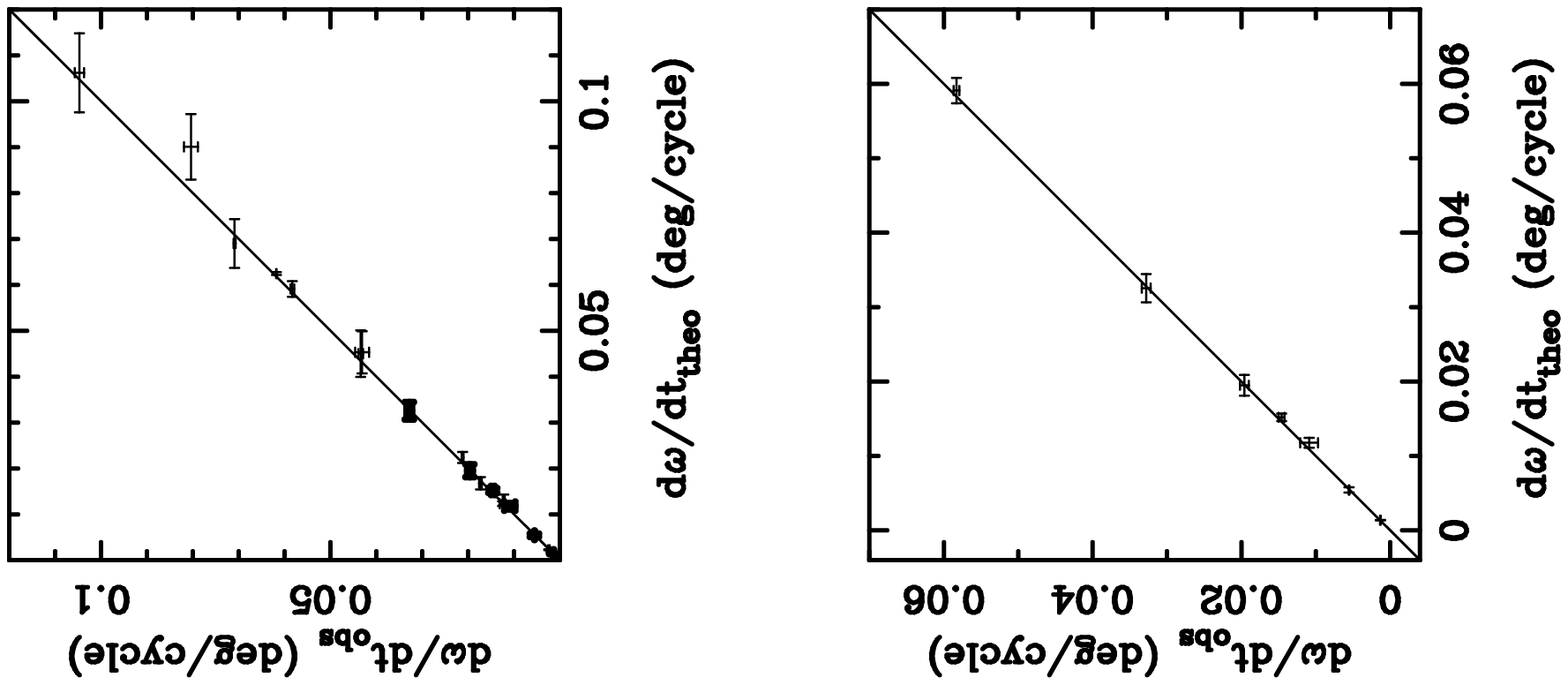}
 \caption{Comparison between $\dot\omega_{\rm theo}$ and $\dot\omega_{\rm obs}$. The top panel shows a comparison between the theoretical predictions and the observed 
 values of the apsidal motion rates for all systems in Table 3 while the bottom panel shows the same comparison but only for systems  with a third body. }
 \label{fig:wdot_comp}
\end{figure}

The comparison of the theoretical $\dot\omega$ with the observed values is shown in Fig.\,\ref{fig:wdot_comp}. 
As can be seen, the agreement between observed and predicted values is excellent. The only system with marginal agreement, at the 
limit of the adopted uncertainties, is EM Car, having the highest mass components of our sample. Such good agreement 
includes the cases where a third body has been identified, as shown in the bottom panel of Figure\,\ref{fig:wdot_comp} (see \ref{subsec:structure_thirdbody}). 

\begin{figure}
 \includegraphics[height=8.cm,width=16cm,angle=-90]{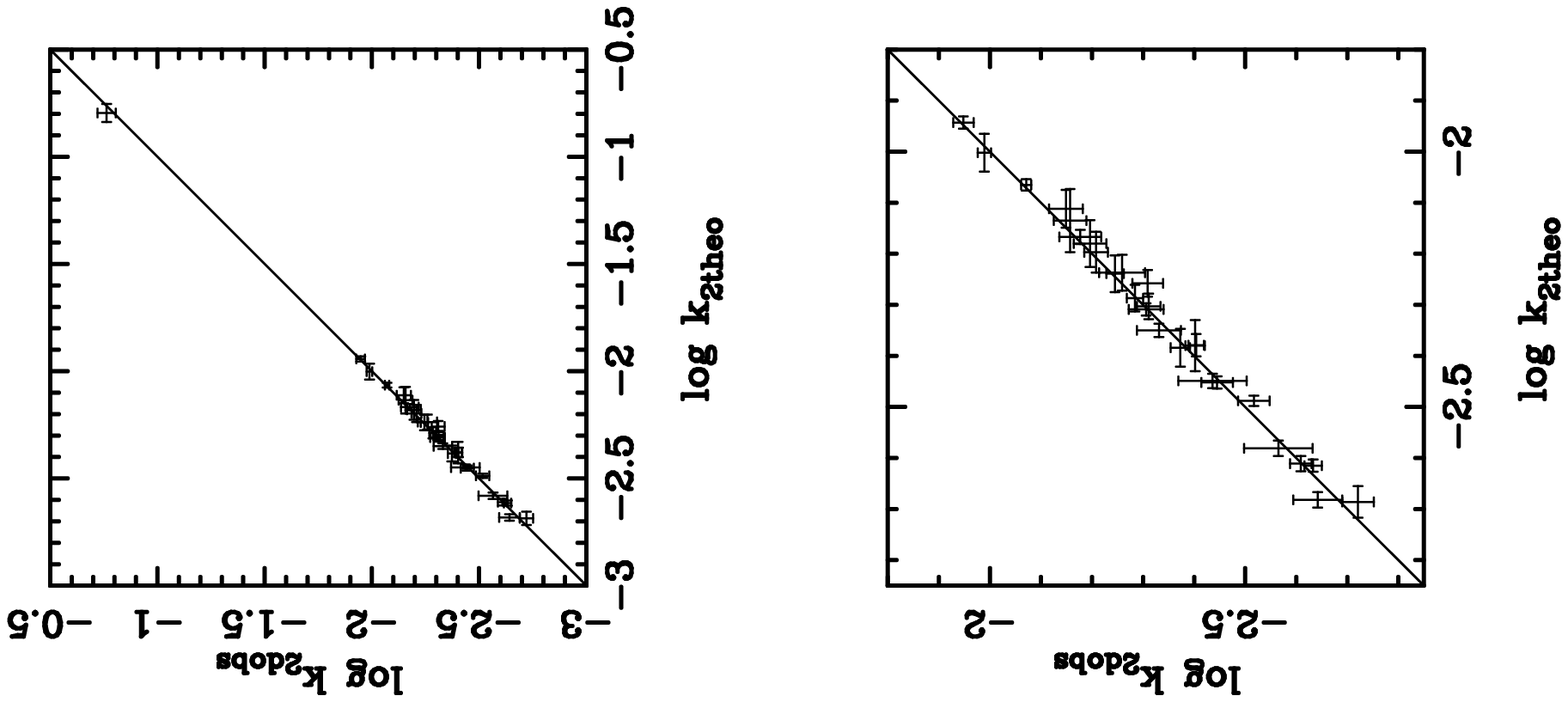}
 \caption{The top panel depicts a comparison between the theoretical predictions and the observed values of the apsidal motion 
constant log k$_2$ for all systems in Table\,\ref{tab:apsidal}. The bottom panel shows the same comparison but without plotting the low mass binary 
V530 Ori, which has a rather extreme value, so that the bulk of the sample can be better visualized.  }
\label{fig:k2_comp}
\end{figure}

For the test of stellar models, we have to use the values of $k_2$. Their observational value is a function of $\dot\omega$ 
and the $c_{2j}$ coefficients given in Eq.\,(\ref{eq:C_weights}):
\begin{equation}
\label{eq:k2_newton}
{\overline k_{2, {\rm obs}}} = { 1 \over {(c_{2,1} + c_{2,2})}} {\dot\omega_{\rm Newt} \over {360}}. 
\end{equation}
\noindent
This refers only to the classical, or Newtonian, term. The total observed apsidal motion rate in Table\,\ref{tab:apsidal} contains the contribution 
of this classical term plus a second additive term corresponding to the relativistic or GR contribution. The GR term is independent 
of the tidal/rotational distortions and can be computed using the equation given by \cite{Levi1937} and rewritten by \cite{Gimenez1985} 
as a function of observable parameters. This equation is, in degrees per cycle:
\begin{equation}
 {\dot\omega_{GR}} = {5.45 \times 10^{-7}} \left({ m_1 + m_2 \over {P}}\right) ^{2/3} {1 \over {(1 - e^2)}},
\end{equation}
\noindent
where the apsidal motion rate corresponding to the relativistic term is expressed in degrees per cycle with masses in solar units 
and the orbital period in days. The classical term in Eq.\,(\ref{eq:k2_newton}), used to calculate the observed $\log k_2$, 
is then simply calculated as
\begin{equation}
 {\dot\omega_{\rm Newt}} = {\dot\omega_{\rm obs}} - {\dot\omega_{\rm GR}}.
\end{equation}

It should be noted that we have assumed the rotational axes of both components to be aligned with the orbital spin in all 
cases except DI Her, for which the expression by \cite{Shakura1985}  was used together with the observed tilt angles 
by \cite{Albrecht2009}. This is further discussed in Sect.\,\ref{subsec:structure_tiltedaxis}, together with the suspected, albeit not corrected, cases of 
V1143 Cyg and EP Cru. Furthermore, all systems for which a third body has been identified through eclipse timing variations
or from the measurement of third light are identified by an asterisk and included in Table\,\ref{tab:apsidal} with no specific correction. 
They are shown in Fig.\,\ref{fig:wdot_comp}, at the bottom, and are discussed in Sect.\,\ref{subsec:structure_thirdbody}.

\begin{figure}
	\includegraphics[height=8.cm,width=6cm,angle=-90]{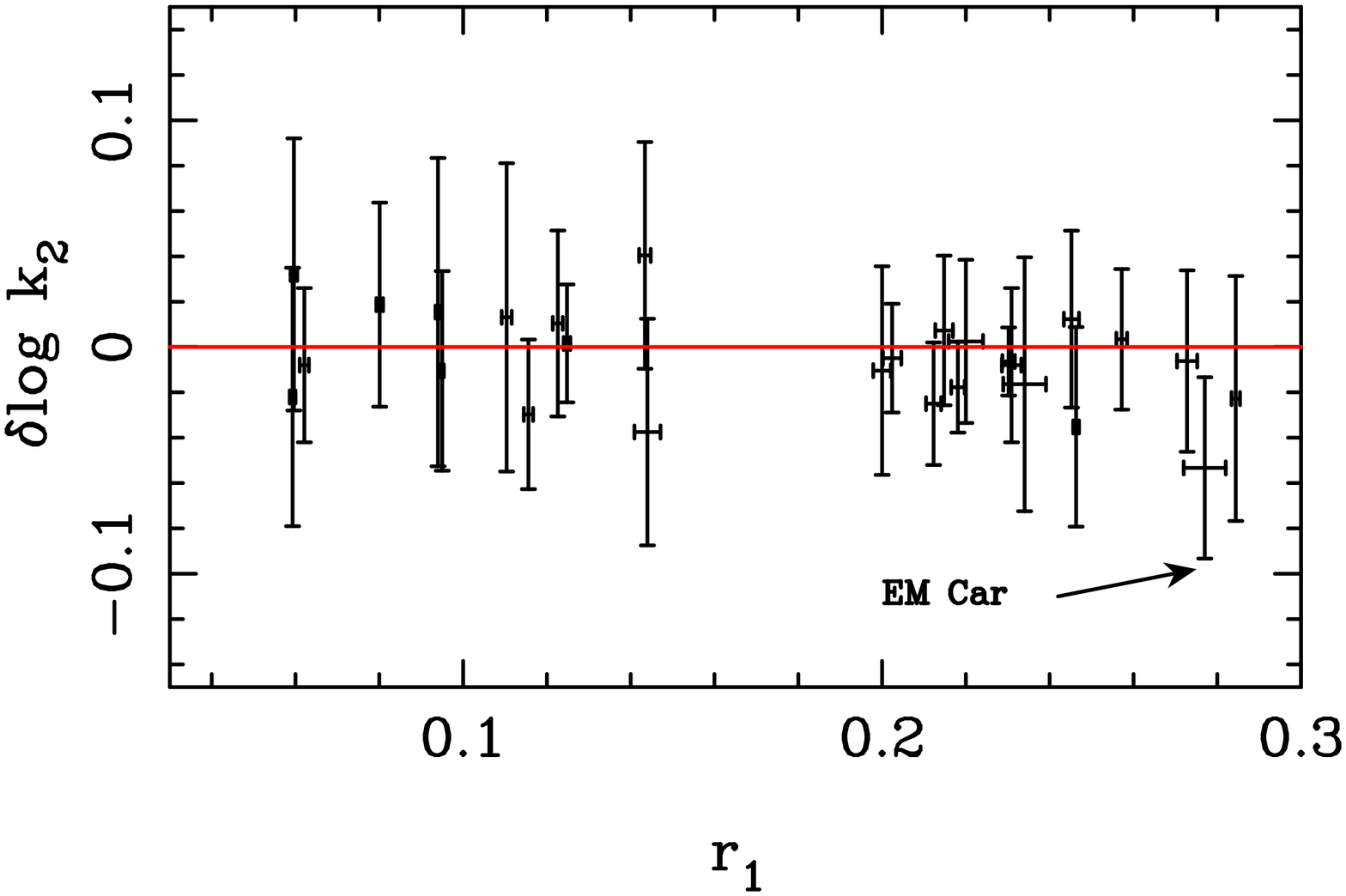}
	\caption{Differences between $\log \overline{k}_{\rm 2obs}$ and $\log \overline{k}_{\rm 2theo}$ as a function of the relative radius of the primary. The 
	mean differential for systems with $r_1< 0.18$ is $\delta \log k_2 = -0.002\pm0.012$ and for those with $r_1>0.18$ is $\delta \log k_2 = -0.010\pm0.008$,  but no significant trend is detected over a constant model.}
	\label{fig:deltalogk2}
\end{figure}

The resulting values of $\log k_2$ derived from the observed apsidal motion rates are given in Table\,\ref{tab:apsidal} and are 
compared with the theoretical values in Fig.\,\ref{fig:k2_comp}. The agreement is very good, with the values of $\log k_2$, 
theoretical and observational, agreeing within their uncertainties for all systems. In spite of the general good agreement, a 
tendency for slightly more concentrated stars (for instance, lower $\log k_2$) than predicted by the models seems 
to be present around $\log k_2=-2.2$, including EM Car. However, we emphasize that all systems agree within the 
corresponding uncertainties. This  effect is not observed in the comparison of $\dot\omega$, shown in Fig.\,\ref{fig:wdot_comp}, except 
for the case of EM Car. A detailed inspection of the observed minus computed values shows that it only affects those binaries with large 
relative radii, above 0.18, and with massive components, the most 
extreme case being indeed EM Car (Fig.\,\ref{fig:deltalogk2}). The deviation might therefore be linked to the observational values used 
for Eq.\,(\ref{eq:C_weights}), given in Table\,\ref{tab:sample}. This could be explained through multiple factors, and given that the deviation is insignificant, we did not attempt to study 
it in further detail in the present work. 

\subsection {Tilted rotational axes} \label{subsec:structure_tiltedaxis}

When the rotational axes of the component stars are tilted with respect to the orbital spin, Eq.\,(\ref{eq:C_weights}) is no longer valid \citep{Company1988} and 
the alternative formulation given by \cite{Shakura1985} is needed. The total Newtonian term of the apsidal motion rate is given by:
 \begin{equation}
  \dot\omega_{\rm Newt} = \dot\omega_{\rm tidal, 1}
  +\dot\omega_{\rm tidal, 2} + \dot\omega_{\rm rot, 1} \phi_1+
  \dot\omega_{\rm rot, 2} \phi_2~,
 \end{equation}
 \noindent where,
 \begin{equation}
  \dot\omega_{{\rm tidal,} j} = \left[{15 m_{3-j}\over{m_{j}}}k_{2j}\thinspace
  g(e)\right]{\left({R_{j}\over{A}}\right)^{5}}~,
  \label{eq:wtidal}
 \end{equation}
 \begin{equation}
  \dot\omega_{{\rm rot,} j} = \left[
  \left({\Omega_{j}\over{\Omega_K}}\right)^{2} \left(1 +
  {m_{3-j}\over{m_{j}}}\right)f(e)\thinspace k_{2j}
  \right]{\left({R_{j}\over{A}}\right)^{5}}~,
  \label{eq:wrot}
 \end{equation}
\begin{eqnarray}
\lefteqn { \phi_j = - {1\over{\sin^2 i}}\left[\cos \alpha_j (\cos \alpha_j -
  \cos \beta_j \cos i)  \right]} \nonumber \\ 
 \lefteqn{-{1\over{2}} \left[(1 - 5\cos^2 \alpha_j)\right].}
\end{eqnarray}
The angle $i$ is the inclination of the orbital plane, $\alpha_j$ are the angles between the rotation axes and the normal to
the orbital plane, and $\beta_j$ are the angles between the rotation axes and the line of sight, while the angles $\lambda_j$ 
are the projection of the spin axes and the orbital axis on the plane of the sky for star $j$. All other parameters are the same 
as in Eq.\,(\ref{eq:C_weights}). The angles $\alpha_j$ and $\beta_j$ cannot be directly measured although they are related by the equation,
 \begin{equation}  
  \cos\alpha_j = \cos\beta_j \cos i + \sin\beta_j \sin i \cos \lambda_j,  
 \end{equation}
 \noindent where the angles $\lambda_j$ can be directly measured using the Rossiter-McLaughlin effect \citep{Rossiter1924,McLaughlin1924}.

The Monte Carlo simulations of apsidal motion rates were performed from realizations of the masses, radii, $k_2$, orbital parameters and  $\lambda_j$ angles considering Gaussian distributions for the error bars. Also, we assumed that $\beta_j$ are distributed randomly, for instance, ${\rm Prob} (\beta_j) d\beta_j = \sin \beta_j d\beta_j$. In other words, $\cos \beta_j$ has a uniform distribution. Finally, using  Eq. (15), the values of $\cos \alpha_j$ can be determined. Angles $\beta_j$ implying equatorial rotational velocities exceeding the breakup velocities were discarded.

\begin{figure}
  \includegraphics[height=8.cm,width=6cm,angle=-90]{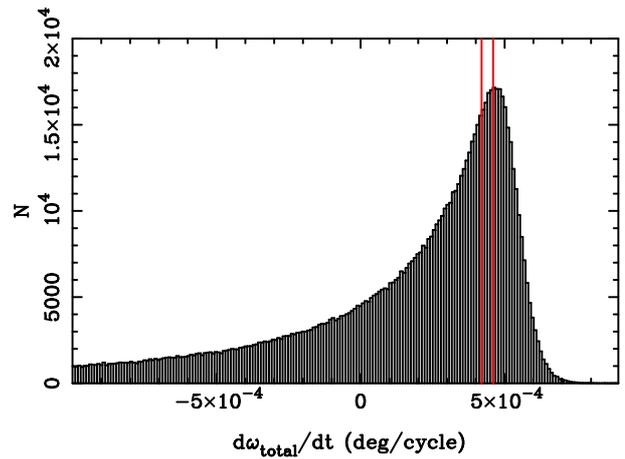}
  \caption{Monte Carlo simulations ($10^6$ realizations) for the apsidal motion of DI Her. The two vertical lines indicate the 
observational error bars.} \label{fig:DIHer_MC}
 \end{figure}

DI Her was puzzling for long time due to the flagrant disagreement between observed and predicted apsidal motion rates \citep{Guinan1985,Claret1998}, but 
 \cite{Albrecht2009} confirmed, using the Rossiter-McLaughlin effect, 
that the rotational axes of the components stars are actually tilted. The authors measured the angles 
$\lambda_1 = +72^{\circ} \pm 4^{\circ}$ and $\lambda_2 = -84^{\circ} \pm 8^{\circ}$ for the primary and the secondary components, 
respectively. Adopting the values of $\log k_2$ derived from the stellar models described in Sect.\,\ref{sec:models} ($\log k_{2, 1}=-2.146\pm0.050$,
$\log k_{2, 2}=-2.171\pm0.050$), the histogram in Fig.\,\ref{fig:DIHer_MC} is generated by simulation of the angles not directly measured. The comparison 
with the observed apsidal motion rate, confirms the solution of the old discrepancy. 

\begin{figure}
  \includegraphics[height=8.cm,width=6cm,angle=-90]{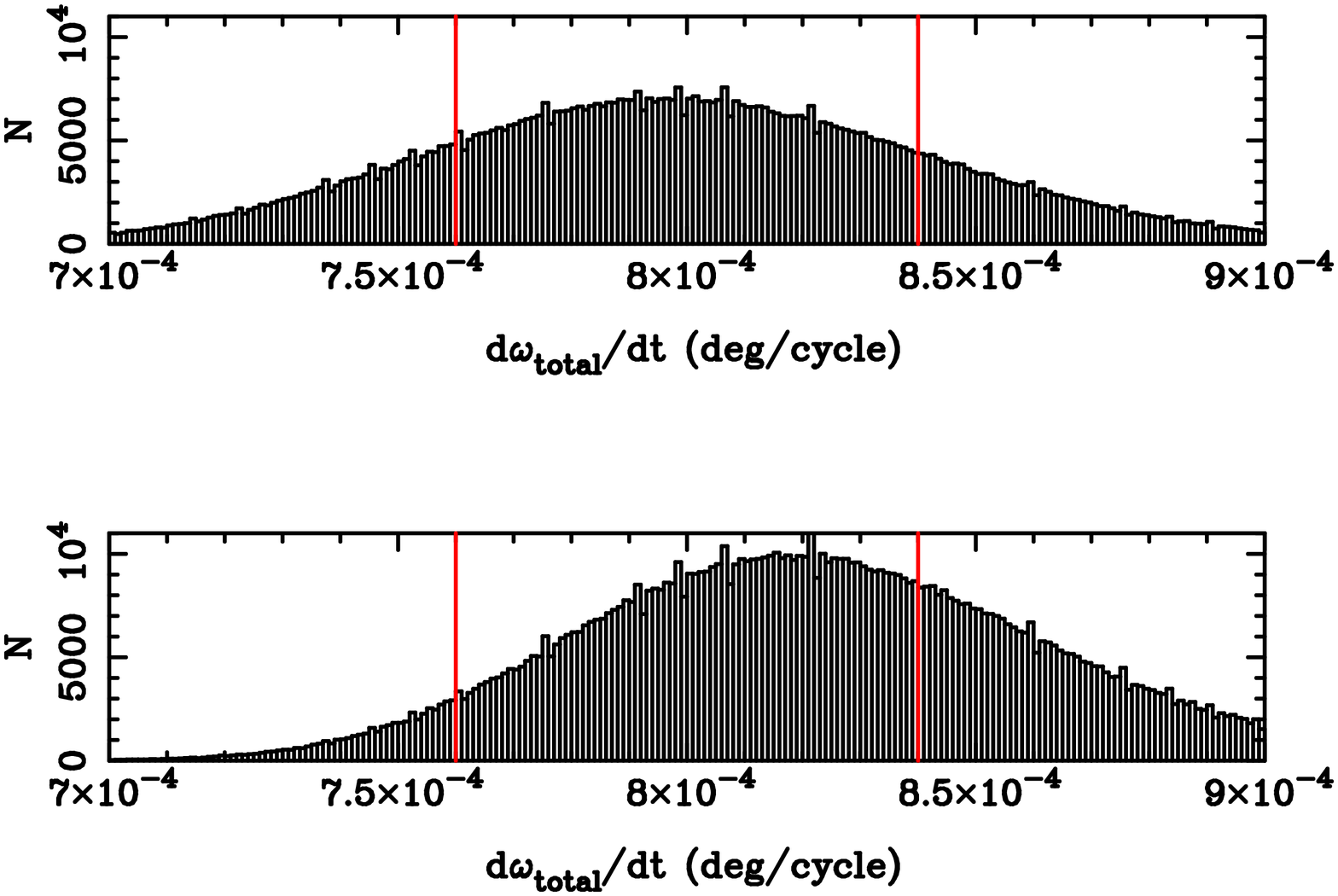}
  \caption{Monte Carlo simulations ($10^6$ realizations) for the total apsidal motion of V1143 Cyg. In the upper panel, we assumed 
misaligment while in the bottom panel we simulate the case of alignment. The two vertical lines indicate the observational error bars. }
\label{fig:V1143Cyg_MC}
 \end{figure}

Two other eclipsing binaries in our sample, V1143 Cyg and EP Cru, have been suspected of having tilted rotational axes. For V1143 Cyg, \cite{Albrecht2007} observed that 
the axes of both components are aligned with the orbital spin within uncertainties. To check the effect in the apsidal motion computations we have 
carried out Monte Carlo simulations for both cases, aligned and misaligned, using their values of $\lambda_1 = 0.3^{\circ} \pm 1.5^{\circ}$ 
and $\lambda_2 = -1.2^{\circ}\pm1.6^{\circ}$. The theoretical internal structure constants derived from our models, namely
log $k_{2, 1}=-2.19\pm0.05$ and $\log k_{2, 2}=-2.29\pm0.05$, were used to generate the histograms shown in Fig.\,\ref{fig:V1143Cyg_MC}. Both simulations 
yield very similar results, as expected from the very small angles $\lambda_{1,2}$ and their uncertainties. Although the agreement 
with the observational value is marginally better in the case of misaligned axes, they both agree within errors and we have adopted the 
aligned solution in Table\,\ref{tab:apsidal}.   

\begin{figure}
  \includegraphics[height=8.cm,width=8cm,angle=-90]{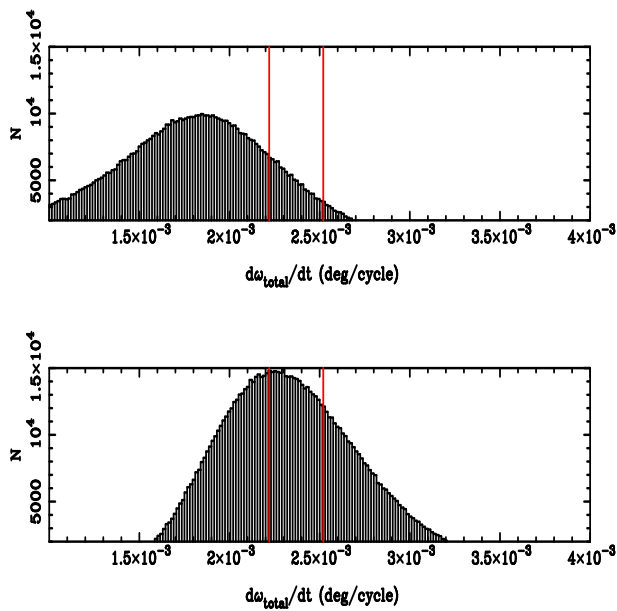}
  \caption{Monte Carlo simulations ($10^6$ realizations) for the total apsidal motion of EP Cru. In the upper panel, we assumed 
misalignment while in the bottom panel we simulate the case of alignment. The two vertical lines indicate the observational error bars.}
\label{fig:EPCru_MC}
 \end{figure}

We have also explored the effect of inclined rotation axes in EP Cru, with direct measurement of the angles 
$\lambda_1 = -1.8^{\circ}\pm 1.6^{\circ}$ and $\left |\lambda_2 \right | < 17^{\circ}$ \citep{Albrecht2013}. 
For the simulation, we adopted the internal structure constants derived from our models as $\log k_{2, 1}=-2.351\pm0.015$ and 
$\log k_{2, 2}=-2.348\pm0.015$, corrected for a significant rotation-induced internal density concentration. The corresponding 
histograms, representing the total predicted apsidal motion, are shown in Fig.\,\ref{fig:EPCru_MC} (upper panel: not aligned; bottom: aligned). The 
solutions based on aligned rotation axes show better agreement with the observed $\dot\omega_{obs}$ and we have adopted such
aligned configuration for Table\,\ref{tab:apsidal}.

\subsection{Systems with a third body} \label{subsec:structure_thirdbody}

We have found that some systems with physically bound third bodies actually perform well in the comparison between the observed
and predicted internal structure constants. To identify possible dynamical effects in the observed values, 
we refer to Fig.\,\ref{fig:wdot_comp}, with the comparison of observed and predicted apsidal motion rates, where systems with a potentially 
perturbing third body are denoted with a red symbol. The good agreement, within the uncertainties, is evident and no systematic 
trend is detected. In fact, slightly faster rates should have been observed, compared to those predicted without considering 
their companions, if an additional term to the total apsidal motion rates is noticeable \citep[see, for instance,][]{Martynov1973}.

For some DLEBs in the sample, the properties of the third body are quite well measured and have determinations
of key parameters such as the orbital period and the mass function. This is the case of CW Cep \citep{Wolf2006}, V539 Ara \citep{Wolf2005b} and 
$\zeta$ Phe \citep{Zasche2007}. For these systems we calculated the additional term in the apsidal motion rate using the approximation 
of \cite{Martynov1973} for coplanar orbits. The resulting corrections were found to be very small compared with the total apsidal motion rate. 
In the case of V539 Ara, that shows the largest contribution, it is on the order of $1.5\times10^{-6}$\,deg/cycle, totally undetectable with errors 
in the observed rate of $0.6\times10^{-3}$\,deg/cycle. Their impact thus should be negligible, as observed.

\subsection {Convective core overshooting}

As a result of the grid search for the best models fitting the observed properties of the component stars in our 
sample, we also derive their values for $f_{\rm ov}$. Unfortunately, our sample is dominated by quite unevolved systems 
that do not provide strong constraints to the core overshooting parameter. We tried, nevertheless, to 
check the consistency of the results with the semi-empirical calibration obtained by \cite{Claret2019} for 
much more evolved binary systems.

\begin{figure}
 \includegraphics[height=8.cm,width=6cm,angle=-90]{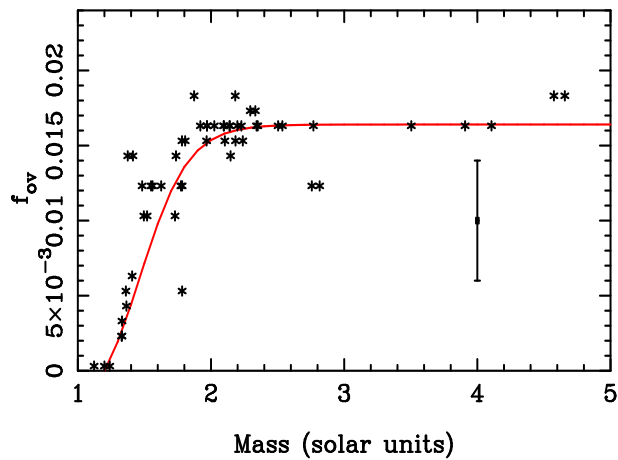}
 \caption{The solid line is an approximate representation of the derived $f_{\rm ov}$ as a function of stellar mass 
obtained by \cite{Claret2019} while the symbols represent the values of $f_{\rm ov}$ resulting from the 
comparison between the theoretical models and the observed absolute dimensions performed in the present study. The error bar 
displayed represents the typical uncertainties for unevolved or mildly evolved stars. 
 For the more massive stars, beyond the calibration of $f_{ov}$  by \cite{Claret2019}, we find an average of the best values from the same methodology to be around 0.025. The only slightly evolved massive system is V453 Cyg, which is discussed in section 5.3, and shows a better fit of the observed parameters with $f_{ov}= 0.03$.}
\label{fig:overshoot_mass}
\end{figure}

We show in Fig.\,\ref{fig:overshoot_mass} the values of $f_{\rm ov}$ resulting from the comparison between the theoretical models and the observed physical 
parameters, together with the relationship between mass and core overshooting obtained by \cite{Claret2019}. The observed correlation 
is quite suggestive, especially when considering that most of our systems have components well within the main sequence. We assume a 
conservative formal error for $f_{\rm ov}$ of 0.005. The most massive stars in our sample cannot be included in such a comparison because they are above the upper limit considered by \cite{Claret2019}, $\approx$5 M$_{\odot}$. 

\begin{figure}
 \includegraphics[height=8.cm,width=6cm,angle=-90]{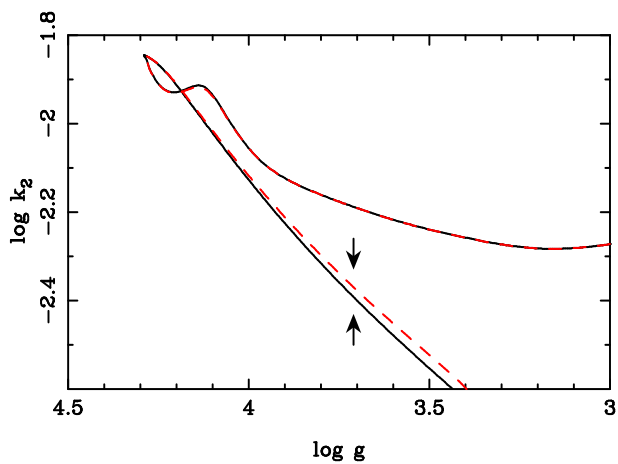}
 \caption{$\log g - \log k_2$ diagram for the primary of V453 Cyg and adopting $Z=0.0134$. The solid line indicates a model with 
$f_{\rm ov} = 0.03$ while the dashed one represents a model computed with $f_{\rm ov} = 0.02$. The two arrows indicate the observed 
value of $\log g$. The resulting difference in $log \overline{\rm k}_{\rm 2,theo}$  is $\approx -0.01$.}
\label{fig:V453Cyg_k2}
\end{figure}

In order to extend the analysis of the variation of $f_{\rm ov}$ to higher masses, we need DLEBs with accurate dimensions like 
those in our sample but, unfortunately, most are quite unevolved systems. At such early evolutionary stages it becomes difficult to 
distinguish the impact of different amounts core overshooting both on the physical parameters and also on log k$_2$. 
The only potential case may be the high-mass system V453 Cyg, whose moderately evolved stage could allow exploring the effect of convective overshooting.
Our methodology to model the DLEB observables, described in Sect. \ref{sec:models} and applied to V453 Cyg, favored $f_{\rm ov} = 0.03$, clearly above the 
highest value given by \cite{Claret2019} for less massive stars. Comparing the corresponding apsidal motion rate with the observed value, $f_{\rm ov} = 0.03$ 
is also favored with respect to adopting 0.02. This is illustrated in Figure\,\ref{fig:V453Cyg_k2}, which shows log k$_2$ for models with 
different amounts of core overshooting ($f_{\rm ov} = 0.02$ -- dashed line; $f_{\rm ov} = 0.03$ -- solid line) for the most massive and evolved component of 
V453 Cyg. We can see that the effect of core overshooting in $\log k_2$ is only noticeable beyond the middle of the main sequence.  
At less evolved stages (including the PMS phase), the differences in the model predictions
when changing the overshooting parameter are well below the observational errors. The differences though become easily observable as we consider 
evolutionary stages approaching the giant branch phase.

The results for V453 Cyg seem to indicate that apsidal motion can help to put constraints on the convective cover overshooting parameter 
using suitable DLEBs with eccentric orbits. This was already explored by \cite{Guinan2000} and revisited by \cite{Claret2003} using the evolved 
high-mass system V380 Cyg.  The question of a dependence of core overshooting with mass using DLEB data has been discussed previously without reaching 
conclusive results \citep{Ribas2000b,Claret2007,Tkachenko2020,Rosu2020}. The tentative increase of $f_{\rm ov}$ with stellar mass that we find in the present 
work would imply a more pronounced dependence than the relationship found by \cite{Claret2019} for stars more massive than $\approx5$ M$_{\odot}$. In the same 
sense, recent theoretical studies carried out by \cite{Martinet2021} also find a need for larger convective cores at higher masses 
\citep[see also][]{Scott2021}. However, the indications found are still not sufficiently well established. Further detailed analyses of suitable massive 
DLEBs should provide the necessary observational evidence to establish a possible overshooting-mass dependence in this interval of masses.

\section {Conclusions} \label{sec:conclusions}

We have studied 34 DLEBs with eccentric orbits employing TESS data that we use to determine eclipse timings. This allowed us to
determine the apsidal motion rates for 27 DLEBs with sufficient precision to compare their internal structure constants with those 
predicted by state-of-the-art theoretical models. The precision of the TESS eclipse timings, as well as the expanded
span of time covered when compared with archival values, have allowed us to significantly improve previous apsidal motion rate 
determinations and detect their presence in some cases for the first time.

The sample was selected in such a way that the dominant term in the apsidal motion is the classical, or Newtonian, contribution. In 
paper I we used a complementary sample to demonstrate that the relativistic term calculated using GR shows excellent agreement with the
observations. Therefore, in the present paper, where we study the internal structure of stars, we subtracted out the relativistic
term to the measured apsidal motion rate to ultimately estimate empirical values for the internal concentration parameters of the 
component stars, namely $\log k_2$. A comparison of our theoretical and observational results reveals excellent agreement.

The case of EM Car deserves further attention. In spite of a reasonable agreement between theory and observation, 
its high mass indicates that some additional effects could be present, 
 including the relative proximity of the components to their Roche limit. For more detailed conclusions, additional observations 
and improved physical parameters are needed. EM Car could also help to eventually understand the small deviations, in the 
same sense, observed in our comparison for massive stars with large relative radii.

A rather unexpected result is the good agreement found for systems with an identified third body. The impact of the 
companion in the dynamical behavior of the close binary should have been observed as systematic differences in the 
observed vs. computed diagram. This is not the case, and probably indicates that the companions are less massive than the 
components of the close system, and/or orbit them at a large distance. For those cases with good enough observational 
constraints on the properties of the third bodies, the negligible contribution of the gravitational effects on the apsidal 
motion,   assuming coplanar orbits,  was confirmed.

The systems in our sample are not especially suitable, owing to their relatively unevolved nature, to provide constraints on the best 
value of the convective core overshooting parameter, $f_{\rm ov}$. Nevertheless, the values resulting from our analysis are clearly 
compatible with the relation found by \cite{Claret2019}. The particular case of V453 Cyg, a high-mass system with moderately 
evolved components, albeit still in the main sequence, indicates a larger value of $f_{\rm ov}$ than that for lower-mass stars. The 
analysis of a larger sample of massive DLEBs will be needed to further constrain the increasing value of the overshooting parameter 
with mass, as suggested in the past in several works.

\longtab{
\footnotesize
\begin{longtable}{lcccccc}
\caption{Astrophysical parameters} \label{tab:sample}\\
  \hline
  \hline
System   & P$_s$ ($d$) & Mass ($M_{\odot}$) & Radius ($R_{\odot}$) & $T_{\rm eff}$ (K)& $v\sin i$ (km\,s$^{-1}$) & Ref. \\   
\hline
 \endfirsthead
\caption{Astrophysical parameters}\\
\hline\hline
System   & P$_s$ ($d$) & Mass ($M_{\odot}$) & Radius ($R_{\odot}$) & $T_{\rm eff}$ (K)& $v\sin i$ (km\,s$^{-1}$) & Ref. \\ 
  \hline
\endhead
\hline
\endfoot 
EM Car   & 3.414281$\pm$0.000005& 22.83$\pm$0.32&  9.35$\pm$0.17& 34000$\pm$2000&  150$\pm$20&1\\
& & 21.38$\pm$0.33&  8.35$\pm$0.16&34000$\pm$2000&  130$\pm$15&\\
Y Cyg    & 2.9963321$\pm$0.0000003& 17.73$\pm$0.30&  5.82$\pm$0.06 &33500$\pm$2000&  147$\pm$10&2\\
& & 17.72$\pm$0.35& 5.79$\pm$0.09& 33200$\pm$2000 & 138$\pm$10&\\
V478 Cyg & 2.880901$\pm$0.000002& 15.40$\pm$0.38&  7.26$\pm$0.09 &30500$\pm$1000&  129.1$\pm$3.6&3\\
& & 15.02$\pm$0.35&  7.15$\pm$0.09&30500$\pm$1000&  127.0$\pm$3.5&\\
V578 Mon & 2.4084822$\pm$0.0000004& 14.54$\pm$0.08&  5.41$\pm$0.04 & 30000$\pm$500&  117$\pm$4 & 4 \\
& & 10.29$\pm$0.06 &  4.29$\pm$0.05 & 25750$\pm$450 &   94$\pm$2 & \\
V453 Cyg & 3.8898249$\pm$0.0000016 & 13.96$\pm$0.23 &  8.670$\pm$0.055 & 28800$\pm$500 &  107.2$\pm$2.8&5\\
& & 11.10$\pm$0.18&  5.250$\pm$0.056 & 27700$\pm$600&   98.3$\pm$3.7&\\
CW Cep   & 2.7291428$\pm$0.0000027 & 12.951$\pm$0.052&  5.520$\pm$0.037 & 28300$\pm$460&  105.2$\pm$2.1&6\\
& & 11.877$\pm$0.049&  5.090$\pm$0.032 & 27430$\pm$430 &   96.2$\pm$1.9 & \\
QX Car   & 4.4779760$\pm$0.0000009&  9.25$\pm$0.12&  4.290$\pm$0.091 &23800$\pm$500&  120$\pm$10 & 1\\
& &  8.46$\pm$0.12&  4.050$\pm$0.091 & 22800$\pm$500&  110$\pm$10&\\
V539 Ara & 3.1690854$\pm$0.0000012&  6.240$\pm$0.066&  4.516$\pm$0.084&18100$\pm$500&   75$\pm$8&1\\
& &  5.314$\pm$0.060&  3.428$\pm$0.083&17100$\pm$500&   48$\pm$5&\\
DI Her   &10.5501696$\pm$0.0000007&  5.17$\pm$0.11&  2.681$\pm$0.046&17000$\pm$800&  108$\pm$10&1\\
& &  4.524$\pm$0.066&  2.478$\pm$0.046&15100$\pm$700&  119$\pm$15&\\
EP Cru   &11.0774707$\pm$0.0000043&  5.02$\pm$0.13&  3.590$\pm$0.035&15700$\pm$500&  141$\pm$5&7\\
& &  4.83$\pm$0.13&  3.495$\pm$0.034&15400$\pm$500&  138$\pm$5&\\
V760 Sco & 1.7309337$\pm$0.0000080&  4.969$\pm$0.090&  3.015$\pm$0.066&16900$\pm$500&   95$\pm$10&1\\
& &  4.609$\pm$0.073&  2.641$\pm$0.053&16300$\pm$500&   85$\pm$10&\\
MU Cas   & 9.652929$\pm$0.000014&  4.657$\pm$0.095&  4.195$\pm$0.058&14750$\pm$800&   22$\pm$2&1\\
& &  4.575$\pm$0.088&  3.670$\pm$0.057&15100$\pm$800&   21$\pm$2&\\
GG Lup   & 1.8495996$\pm$0.0000015&  4.106$\pm$0.044&  2.380$\pm$0.025&14750$\pm$450&   97$\pm$8&1\\
& &  2.504$\pm$0.023&  1.726$\pm$0.019&11000$\pm$600&   61$\pm$5&\\
$\zeta$ Phe & 1.669774$\pm$0.000026&  3.908$\pm$0.057&  2.835$\pm$0.019 & 14400$\pm$800&   85$\pm$8 & 16 \\
& &  2.536$\pm$0.031&  1.885$\pm$0.011 & 12000$\pm$600&   75$\pm$8&\\
IQ Per   & 1.7435699$\pm$0.0000001&  3.504$\pm$0.054&  2.445$\pm$0.024&12300$\pm$230&   68$\pm$2&1\\
& &  1.730$\pm$0.025&  1.499$\pm$0.016& 7700$\pm$140&   44$\pm$2&\\
PV Cas   & 1.7504698$\pm$0.0000009&  2.816$\pm$0.050&  2.301$\pm$0.020&10200$\pm$250&   67$\pm$5 & 1\\
& &  2.757$\pm$0.054&  2.257$\pm$0.019&10190$\pm$250&   66$\pm$5 &\\
V364 Lac & 7.3515458$\pm$0.0000043&  2.333$\pm$0.014&  3.309$\pm$0.021& 8250$\pm$150&   45$\pm$1&1\\
& &  2.295$\pm$0.024&  2.986$\pm$0.020& 8500$\pm$150&   15$\pm$1&\\
SW CMa   &10.091978$\pm$0.000005&  2.239$\pm$0.014&  3.014$\pm$0.020& 8200$\pm$150&   24.0$\pm$1.5&8\\
& &  2.104$\pm$0.018&  2.495$\pm$0.042& 8100$\pm$150&   10.0$\pm$1.0&\\
PT Vel   & 1.8020075$\pm$0.0000010&  2.198$\pm$0.016&  2.094$\pm$0.020& 9250$\pm$150&   63$\pm$2&9\\
& &  1.626$\pm$0.009&  1.559$\pm$0.020& 7650$\pm$150&   40$\pm$3&\\
V1647 Sgr& 3.2827992$\pm$0.0000050&  2.184$\pm$0.037&  1.832$\pm$0.018& 9600$\pm$300&   80$\pm$5&1\\
& &  1.967$\pm$0.033&  1.667$\pm$0.017& 9100$\pm$300&   70$\pm$5&\\
AI Hya   & 8.2896499$\pm$0.0000012&  2.140$\pm$0.030&  3.960$\pm$0.040& 6860$\pm$200&   40$\pm$10 &11\\
& &  1.970$\pm$0.030&  2.810$\pm$0.020& 7290$\pm$200&   29$\pm$10 &\\
VV Pyx   & 4.5961832$\pm$0.0000050&  2.097$\pm$0.022&  2.168$\pm$0.020& 9500$\pm$200&   23$\pm$3&1\\
& &  2.095$\pm$0.019&  2.168$\pm$0.020& 9500$\pm$200&   23$\pm$3&\\
EK Cep   & 4.4277960$\pm$0.0000003&  2.025$\pm$0.023&  1.580$\pm$0.007& 9000$\pm$200&   23$\pm$2&1\\
& &  1.122$\pm$0.012&  1.315$\pm$0.006& 5700$\pm$200&   10.5$\pm$2.0&\\
VV Crv   & 3.1445358$\pm$0.0000097&  1.978$\pm$0.010&  3.375$\pm$0.010& 6500$\pm$200&   81$\pm$3&12\\
& &  1.513$\pm$0.008&  1.650$\pm$0.008& 6640$\pm$200&   24$\pm$2&\\
IM Per   & 2.2542269$\pm$0.0000002&  1.7831$\pm$0.0094&  2.409$\pm$0.018& 7580$\pm$150&   57.5$\pm$3.0&13\\
& &  1.7741$\pm$0.0097&  2.366$\pm$0.017& 7570$\pm$160&   56.5$\pm$3.0&\\
BP Vul   & 1.9403466$\pm$0.0000017&  1.737$\pm$0.015&  1.852$\pm$0.014& 7715$\pm$150&   45.4$\pm$0.9&1\\
& &  1.4081$\pm$0.0087&  1.489$\pm$0.013& 6810$\pm$150&   40.5$\pm$1.4&\\
V1022 Cas&12.1561598$\pm$0.0000008&  1.6263$\pm$0.0011&  2.591$\pm$0.026& 6450$\pm$120&   10.9$\pm$1.2&16\\
& &  1.6086$\pm$0.0012&  2.472$\pm$0.027& 6590$\pm$110&    7.0$\pm$1.3&\\
PV Pup   & 1.660728$\pm$0.000003&  1.561$\pm$0.011&  1.543$\pm$0.016& 6920$\pm$300&   43$\pm$4&1\\
& &  1.550$\pm$0.013&  1.499$\pm$0.016& 6930$\pm$300&   43$\pm$4&\\
BF Dra   &11.2110011$\pm$0.0000020&  1.414$\pm$0.003&  2.086$\pm$0.012& 6360$\pm$150&   10.5$\pm$1.8&10\\
& &  1.375$\pm$0.003&  1.922$\pm$0.012& 6400$\pm$150&    9.0$\pm$1.8&\\
V1143 Cyg& 7.640735$\pm$0.000004&  1.361$\pm$0.004&  1.348$\pm$0.016& 6470$\pm$100&   19.6$\pm$0.1&14\\
& &  1.332$\pm$0.004&  1.322$\pm$0.018& 6470$\pm$100&   28.2$\pm$0.1&\\
IT Cas   & 3.896649$\pm$0.000025&  1.3315$\pm$0.0093&  1.594$\pm$0.018& 6470$\pm$100&   19$\pm$2&1\\
& &  1.3290$\pm$0.0078&  1.562$\pm$0.040& 6470$\pm$100&   17$\pm$2&\\
AI Phe   &24.59215$\pm$0.00002&  1.2438$\pm$0.0007&  2.907$\pm$0.013& 5240$\pm$150&    6$\pm$1&15\\
& &  1.1941$\pm$0.0007&  1.841$\pm$0.017& 6310$\pm$150&    4$\pm$1&\\
\newpage
EW Ori   & 6.9368442$\pm$0.0000004&  1.173$\pm$0.011&  1.168$\pm$0.005& 6070$\pm$100&    9.0$\pm$0.7&17\\
& &  1.123$\pm$0.009&  1.097$\pm$0.005& 5900$\pm$100&    8.8$\pm$0.6&\\
V530 Ori & 6.1107784$\pm$0.0000003&  1.0038$\pm$0.0066&  0.980$\pm$0.013& 5890$\pm$100&    9$\pm$1&18\\
& &  0.5955$\pm$0.0022&  0.587$\pm$0.007& 3880$\pm$120&    5$\pm$1 &\\
\hline
\end{longtable}

\footnote { References to Table 1:
 {\bf 1}  \cite{Torres2010};
 {\bf 2} \cite{Harmanec2018};
 {\bf 3}  \cite{Pavlovski2018};
 {\bf 4} \cite{Garcia2014};
 {\bf 5}  \cite{Southworth2020a};
 {\bf 6}  \cite{Lee2021};
 {\bf 7}  \cite{Albrecht2013};
 {\bf 8}  \cite{Torres2012};
 {\bf 9}  \cite{Bakis2008};
 {\bf 10} \cite{Lacy2012b};
 {\bf 11} \cite{Lee2020};
 {\bf 12} \cite{Fekel2013};
 {\bf 13} \cite{Lacy2015};
 {\bf 14} \cite{Lester2019};
 {\bf 15} \cite{Maxted2020};
 {\bf 16} \cite{Southworth2020b};  
 {\bf 17} \cite{Clausen2010};   
 {\bf 18} \cite{Torres2014}. 
}}

\renewcommand{\tablename}{Table }   

\begin{table}[t]
\centering
\caption{$T_2-T_1$ computed from TESS lightcurves. } \label{tab:t2t1}
\begin{tabular}[t]{lccc}
\hline
\hline
\noalign{\smallskip}
System  \& $T_0$  & $N$  & $T_2-T_1$ [d] & $dN$  \\
\noalign{\smallskip}
\hline
\noalign{\smallskip}
EM Car       &    0 & $ 1.723492\pm0.000045$ & 0\\
2458572.9908 &    1 & $ 1.723193\pm0.000101$ & 0\\
     &           2 &  $ 1.723907\pm0.000082$ & 0 \\
     &           4 &  $ 1.724324\pm0.000080$ & 0 \\
     &           5 &  $ 1.723740\pm0.000054$ & 0 \\
     &           6 &  $ 1.724186\pm0.000064$ & 0 \\
     &           8 &  $ 1.723906\pm0.000098$ & 0 \\
     &           9 &  $ 1.723282\pm0.000064$ & 0 \\
     &          10 &  $ 1.722915\pm0.000061$ & 0 \\
     &          12 &  $ 1.724228\pm0.000082$ & 0 \\
     &          13 &  $ 1.722725\pm0.000064$ & 0 \\
     &          14 &  $ 1.724169\pm0.000084$ & 0 \\
\hline
Y Cyg        &    0 & $ 1.743685\pm0.000058$ & 0\\
2458713.0054 &    1 & $ 1.743724\pm0.000035$ & 0\\
     &           2 &  $ 1.743574\pm0.000034$ & -1 \\
     &           5 &  $ 1.743077\pm0.000032$ & 0 \\
     &           6 &  $ 1.742991\pm0.000025$ & 0 \\
     &           7 &  $ 1.742894\pm0.000027$ & -1 \\
\hline
\end{tabular}
\tablefoot{The BJD value below each system name defines the origin epoch 
	of the orbital cycle count ($N$). This table is presented in its entirety in the online version of the article.}
\end{table}

\renewcommand{\tablename}{Table }       
\begin{table*}  
\caption{Observed and theoretical apsidal motion rates} \label{tab:apsidal}%
\begin{flushleft}
 \begin{tabular}{lcccccccc}    
  \hline
  \hline
   System        & Method** & $e$ & $\dot\omega_{\rm theo}$ &$\dot\omega_{\rm obs}$ & $\log \overline{k}_{\rm 2,theo}$ & $\log \overline{k}_{\rm 2,obs}$    & \\
  \hline  
  EM Car        &B &0.0120$\pm$0.0005 &0.090$\pm$0.007 &0.080$\pm$0.002& $-$2.258$\pm$0.026 & $-$2.311$\pm$0.030&\\
  Y Cyg         &L &0.14508$\pm$0.00029 &0.06244$\pm$0.00031   &0.06178$\pm$0.00003& $-$1.943$\pm$0.012 & $-$1.948$\pm$0.020&\\
  V478 Cyg      &L &0.0158$\pm$0.0007 &0.1062$\pm$0.0086   &0.1047$\pm$0.0010& $-$2.239$\pm$0.036 & $-$2.245$\pm$0.017&\\
  V578 Mon      &L &0.07755$\pm$0.00026 &0.0690$\pm$0.0053   &0.07089$\pm$0.00021& $-$2.002$\pm$0.037 & $-$1.989$\pm$0.013&\\
  V453 Cyg      &B &0.0250$\pm$0.0014 &0.0453$\pm$0.0046   &0.0431$\pm$0.0015& $-$2.380$\pm$0.050 & $-$2.402$\pm$0.019&\\
  CW Cep*       &B &0.0305$\pm$0.0009 &0.0591$\pm$0.0017   &0.0583$\pm$0.0004& $-$2.065$\pm$0.011 & $-$2.071$\pm$0.010&\\
  QX Car        &L &0.278$\pm$0.003 &0.0132$\pm$0.0011   &0.01222$\pm$0.00022& $-$2.112$\pm$0.037 & $-$2.149$\pm$0.033&\\
  V539 Ara*     &B &0.0530$\pm$0.0010 &0.0195$\pm$0.0014   &0.0196$\pm$0.0006& $-$2.309$\pm$0.012 & $-$2.306$\pm$0.034&\\
  DI Her        &A &0.489$\pm$0.003 &0.00046$\pm$0.00009   &0.00044$\pm$0.00002& $-$2.135$\pm$0.062 & $-$2.157$\pm$0.032&\\
  EP Cru        &A &0.1874$\pm$0.0005 &0.00229$\pm$0.00011   &0.00237$\pm$0.00015& $-$2.350$\pm$0.013 & $-$2.331$\pm$0.043&\\
  V760 Sco      &B &0.0265$\pm$0.0010 &0.0450$\pm$0.0051   &0.0434$\pm$0.0005& $-$2.180$\pm$0.046 & $-$2.196$\pm$0.032&\\
  GG Lup        &B &0.155$\pm$0.005 &0.0168$\pm$0.0014   &0.0173$\pm$0.0003& $-$2.197$\pm$0.040 & $-$2.208$\pm$0.023&\\
  $\zeta$ Phe*  &B &0.0116$\pm$0.0024 &0.0326$\pm$0.0019   &0.0328$\pm$0.0006& $-$2.287$\pm$0.026 & $-$2.284$\pm$0.016&\\
  IQ Per        &B &0.0662$\pm$0.0005 &0.01526$\pm$0.00088   &0.0150$\pm$0.0005& $-$2.303$\pm$0.025 & $-$2.311$\pm$0.023&\\
  PV Cas        &B &0.0325$\pm$0.0005 &0.0224$\pm$0.0012   &0.0212$\pm$0.0002& $-$2.379$\pm$0.022 & $-$2.404$\pm$0.015&\\
  V364 Lac      &A &0.2873$\pm$0.0014 &0.00180$\pm$0.00005   &0.00181$\pm$0.00006& $-$2.611$\pm$0.015 & $-$2.609$\pm$0.021&\\
  SW CMa        &A &0.3180$\pm$0.0005 &0.00067$\pm$0.00002   &0.00069$\pm$0.00005& $-$2.581$\pm$0.015 & $-$2.565$\pm$0.067&\\
  PT Vel        &B &0.112$\pm$0.003 &0.01231$\pm$0.00049   &0.0125$\pm$0.0006& $-$2.452$\pm$0.012 & $-$2.445$\pm$0.031&\\
  V1647 Sgr*    &B &0.4130$\pm$0.0005 &0.00543$\pm$0.00036   &0.00554$\pm$0.00005& $-$2.384$\pm$0.037 & $-$2.373$\pm$0.019&\\
  AI Hya        &B &0.234$\pm$0.002 &0.00177$\pm$0.00015   &0.00191$\pm$0.00005& $-$2.682$\pm$0.015 & $-$2.642$\pm$0.048&\\
  VV Pyx*       &A &0.0956$\pm$0.0009 &0.00138$\pm$0.00004   &0.00132$\pm$0.00005& $-$2.488$\pm$0.010 & $-$2.517$\pm$0.031&\\
  EK Cep        &A &0.109$\pm$0.003 &0.00089$\pm$0.00002   &0.00088$\pm$0.00004& $-$2.167$\pm$0.014 & $-$2.177$\pm$0.041&\\
  VV Crv*       &B &0.0852$\pm$0.0010 &0.01177$\pm$0.00068   &0.0109$\pm$0.0012& $-$2.686$\pm$0.031 & $-$2.721$\pm$0.031&\\
  IM Per*       &B &0.0491$\pm$0.0010 &0.01520$\pm$0.00050   &0.0146$\pm$0.0004& $-$2.615$\pm$0.012 & $-$2.633$\pm$0.017&\\
  V1143 Cyg     &A &0.5386$\pm$0.0012 &0.00082$\pm$0.00003   &0.00080$\pm$0.00004& $-$2.237$\pm$0.035 & $-$2.259$\pm$0.045&\\
  IT Cas        &A &0.089$\pm$0.002 &0.00112$\pm$0.00005   &0.00114$\pm$0.00010& $-$2.449$\pm$0.014 & $-$2.436$\pm$0.067&\\
  V530 Ori      &A &0.0862$\pm$0.0010 &0.00081$\pm$0.00006   &0.00086$\pm$0.00005& $-$0.796$\pm$0.042 & $-$0.763$\pm$0.043&\\  
  
  \hline
 \end{tabular}
 \footnotesize{ The values of  $\dot\omega$ are expressed in deg/cycle.} 
  
 \footnotesize{* Binary systems with a suggested third body.}
 
 \footnotesize{** Adopted methodology to compute $\dot\omega_{\rm obs}$: A, from linear variation of $T_{2}-T_{1}$; B, from variations of the angle of periastron; L, adopted from the literature.}
\end{flushleft}
\end{table*}

\begin{acknowledgements} 
This paper is dedicated to the memory of J. Andersen, a very good 
friend and colleague, that initiated a systematic approach to the accurate measurement of 
masses and radii in DLEBs, and motivated us to continue his path.  We  thank an anonymous 
referee for the insightful  comments and suggestions. 
The Spanish MEC (ESP2017-87676-C5-2-R,  PID2019-107061GB-C64, and  
PID2019-109522GB-C52) is gratefully acknowledged for its 
support during the development of this work. A.C. also 
acknowledges financial support from the State Agency for 
Research of the Spanish MCIU through the “Center of 
Excellence Severo Ochoa” award for the Instituto de 
Astrofísica de Andalucía (SEV-2017-0709). D.B., I.R., and J.C.M. acknowledge
support from the Spanish Ministry of Science and Innovation and the European 
Regional Development Fund through grant PGC2018-098153-B-C33, 
from the Generalitat de Catalunya/CERCA programme, and from the Agència de
Gestió d’Ajuts Universitaris i de Recerca of the Generalitat de Catalunya.
This paper includes data collected by the TESS mission. Funding 
for the TESS mission is provided by the NASA Explorer Program. 
This research has made use of the SIMBAD database, operated at the CDS, Strasbourg, 
France, and of NASA's Astrophysics Data System Abstract Service.
\end{acknowledgements}

\bibliographystyle{aa} 
\bibliography{bibtex.bib}

\begin{appendix} 

 \section{Apsidal motion studies}
 
  \begin{itemize}
  	
  \item EM Car
  
  \cite{Andersen1989} made a complete study of EM Car obtaining an eccentricity $e=0.0120\pm0.0005$ and an apsidal motion rate 
  $\dot\omega=0.081\pm0.010$ deg/cycle. This value has a relatively large error bar because of the unfavorable configuration 
  with $\omega \approx 0$ deg, which affects the use of a linear fit to the differential timings. The light curve solution   
  by \cite{Andersen1989} yields    an argument of periastron of $\omega = 350\pm5$ deg, and the TESS measurements give a phase 
  for the secondary eclipse of 0.50484$\pm$0.00010 which corresponds, adopting also $e = 0.0120\pm0.0005$, to a value of $\omega = 308\pm2$ deg. We used the
  TESS light curve to exclude the other possible value with positive $e sin\omega$. By using the two $\omega$ determinations at
  different epochs we obtain an apsidal motion rate $\dot\omega = 0.080\pm0.002$ deg/cycle, considering all uncertainties. This value is
  in excellent agreement, but significantly more precise, than that given by \cite{Andersen1989} thanks to the large time-span increase, 
  which is now covering nearly a complete apsidal motion period. Furthermore, \cite{Mayer2004} measured $T_2-T_1 = 1.683\pm0.003$ days 
  at HJD 2452280, between the two light curve solutions. This is in excellent agreement with the predicted value from our apsidal 
  motion solution $T_2-T_1 = 1.6828\pm0.0005$ days.
  
  \item  Y Cyg
  
  The physical parameters of Y Cyg were given in the comprehensive study of  \cite{Harmanec2014}, who analyzed 
  light curve changes with time, using a variable $\omega$, and obtained an orbital eccentricity $e=0.1451\pm0.0003$ 
  together with a precise apsidal motion rate of $\dot\omega=0.06178\pm0.00004$ deg/cycle. The authors also studied the changing 
  position of the times of eclipse, obtaining a less precise $e = 0.1448\pm0.0012$, using old photographic timings.
  An earlier study with the same methodology but restricted to the best photoelectric measurements was carried out by 
  \cite{Holmgren1995}, determining $e = 0.1458\pm0.0007$.  
  Our TESS data, given in Table \ref{tab:t2t1}, indicate a clear apsidal motion variation in the observed $T_2-T_1$, with a 
  slope of $-0.000138\pm0.000009$ d/cycle and a value of $T_2-T_1 = 1.74357\pm0.00003$ days at BJD 2458718.9981. 
  The apsidal motion rate cannot be accurately determined using these data alone due to the very narrow time interval, but
  a value of $\dot\omega = 0.062\pm0.004$, with $\omega = 27.8\pm0.3$ deg, is obtained assuming 0.1451 for the eccentricity. 
  Nevertheless, it confirms the solution by \citep{Harmanec2014}, based on the  global analysis of all photometric data with
  variable longitude of the periastron, that we adopt for our discussion.
  
  \item  V478 Cyg
  
  The absolute dimensions of V478 Cyg were recently measured by \cite{Pavlovski2018} using their own radial velocity 
  curves and a new analysis of the light curve of \cite{Sezer1983}. Unfortunately the light curve could not establish a 
  precise value of the orbital eccentricity and $e = 0.021\pm0.005$, as derived from the radial velocity curve, 
  was fixed for the analysis. The analysis of the TESS light curve using the Wilson-Devinney code \citep{Wilson1971}
  resulted in a best fit with $e = 0.016$. This is in agreement with the apsidal motion variation produced by the eclipse timings presented 
  by \cite{Wolf2006} yielding $e = 0.0158\pm0.0007$ and $\dot\omega = 0.1047\pm0.0010$ deg/cycle. TESS 
  measurements give an average $T_2-T_1$ value of $1.4120\pm0.0005$ days, in excellent agreement with the expected 
  value, $T_2-T_1 = 1.4122\pm0.0005$ days, using the solution by \cite{Wolf2006}. The comparison of our TESS 
  data with the $T_2-T_1$ extracted from Table 1 of \cite{Wolf2006}, with timings within less than 10 orbital cycles, 
  confirms the apsidal motion rate determination by \cite{Wolf2006}. However, using the larger and less precise values of the 
  eccentricity $e=0.019\pm0.002$ by \cite{Mossakovskaya1996} or $e=0.021\pm0.005$ by \cite{Pavlovski2018} produces much poorer 
  fits to the data and predictions that are in contradiction with the TESS light curve. We therefore adopted the solution 
  given by \cite{Wolf2006} for Table\,\ref{tab:apsidal}.
  
  \item  V578 Mon
  
  The most recent and complete determination of the physical parameters of V578 Mon was carried out by \cite{Garcia2014} 
  adopting an earlier apsidal motion determination \citep{Garcia2011} of $\dot\omega = 0.07089\pm0.00021$ deg/cycle
  with $e = 0.07755\pm0.00026$. Their global analysis of the light curve variations with $\dot\omega$ as a 
  free parameter over one complete apsidal motion period allowed such precise results. The TESS measurements 
  of the $T_2-T_1$ differences, obtained in two epochs, yield a less precise determination $\dot\omega = 0.0697\pm0.0010$ deg/cycle, 
  assuming the eccentricity of \cite{Garcia2011}. The individual light curve solutions in table 3 by \cite{Garcia2011} give 
  an apsidal motion rate $\dot\omega = 0.0718\pm0.0012$ deg/cycle, and including the arguments of periastron derived from the TESS 
  $T_2-T_1$ values, yield a rate of $\dot\omega = 0.0701\pm0.0006$ deg/cycle. All values agree within their uncertainties and we adopted 
  for our discussion $\dot\omega = 0.07089\pm0.00021$ deg/cycle, as derived from the light curve analysis with variable omega.
  
  \item  V453 Cyg
  
  A detailed analysis of the TESS light curve of V453 Cyg has been recently carried out by \cite{Southworth2020a} 
  in their quest to use $\beta$ Cep pulsations as a tracer of the physical processes that govern the evolution of massive stars.
  The authors determined the orbital eccentricity to be $e = 0.0250\pm0.0014$ with the argument of periastron at 
  $\omega = 152.5\pm5.1$ deg, but did not analyze the apsidal motion variations. This was done in an earlier paper \citep{Southworth2004} 
  with less precise data, providing an apsidal motion rate $\dot\omega = 0.0577\pm0.0016$ deg/cycle, with an orbital eccentricity 
  $e = 0.022\pm0.003$. Nevertheless, this solution predicts an argument of periastron at the time of the TESS observations that is
  not compatible with the TESS light curve analyzed by \cite{Southworth2020a}. We have used 
  the TESS measurements in Table \ref{tab:t2t1}, giving an average eclipse timing difference $T_2-T_1 = 1.8896\pm0.00015$ days, and the timings
  by \cite{Wachmann1973} and \cite{Cohen1971}, using only photoelectric measurements separated by less than 10 orbital cycles. We computed the 
  corresponding arguments of periastron, assuming $e = 0.0250\pm0.0014$, and a linear fit yielded 
  $\dot\omega = 0.0431\pm0.0015$ deg/cycle, considering the uncertainty of the eccentricity.
  
  \item  CW Cep
  
  The light curve of this eclipsing binary system has been recently studied by \cite{Lee2021} using TESS measurements. 
  An eccentricity $e=0.0305\pm0.0009$, with an argument of periastron $\omega = 212.7\pm3.2$ deg were obtained from the analysis. 
  The authors do not discuss the apsidal motion rate, well known from previous studies, but focus on the analysis of the detected 
  $\beta$ Cep pulsations and confirm the presence of a significant third light. The physical parameters of the components are 
  well determined using the radial velocity amplitudes measured by \cite{Johnston2019}. The apsidal motion of CW Cep was 
  studied most recently by \cite{Wolf2006}, who clearly observed the effect of the third body through the light-time effect, 
  with a period of $P_3 = 38.5\pm1.5$ years. After correction for the third-body effect in the eclipse timings, the authors could 
  determine an orbital eccentricity $e=0.0297\pm0.0005$ and an apsidal motion rate $\dot\omega = 0.0582\pm0.0005$ deg/cycle. 
  With the TESS measurements of $T_2-T_1$ in different sectors, apsidal motion is already evident and we determine a slope of 
  $(2.8\pm0.2) \times 10^{-5}$ days/cycle over a time span of 70 orbital cycles. We have added the $T_2-T_1$ values retrieved from 
  the literature to compute the corresponding argument of periastron with the orbital eccentricity given by \cite{Lee2021}. We have 
  restricted eclipse timings to those obtained by means of photoelectric measurements and considered pairs within less than 10 orbital 
  cycles. A weighted linear fit yields $\dot\omega = 0.0583\pm0.0004$ deg/cycle, in excellent agreement with previous determinations.
  
  \item QX Car
  
  Precise absolute dimensions of QX Car were obtained by \cite{Andersen1983}. The authors also studied apsidal motion, fixing the 
  eccentricity to the result of the light curve analysis, $e = 0.278\pm0.003$, and obtained $\dot\omega = 0.01248\pm0.00015$ deg/cycle. 
  A revised analysis by \cite{Gimenez1986}, with the same adopted eccentricity, yielded $\dot\omega = 0.01222\pm0.00022$ deg/cycle 
  considering more realistic uncertainties. TESS measurements give an average value of $T_2-T_1 = 1.4853\pm0.0002$ days. Adopting the 
  eccentricity by \cite{Andersen1983}, this time difference yields an argument of periastron of $\omega=163.1\pm2.5$ deg. The apsidal solution
  by \cite{Gimenez1986} predicts a more precise value of $\omega=164.6\pm0.8$ deg at the time of the TESS observations, in good agreement with 
  the value obtained from the $T_2-T_1$ measurements, and we have adopted it for our discussion. 
  
  \item  V539 Ara
  
  Accurate physical parameters of this system were obtained by \cite{Clausen1996}, who also detected apsidal motion 
  with $\dot\omega = 0.021\pm0.002$ deg/cycle but showing significant variations between different epochs that suggested the 
  presence of a perturbing third body. No third light was reported. The light curve provided a value $e = 0.053\pm0.001$.
  The analysis by \cite{Wolf2005} confirmed the presence of a third body by means of the light-time effect, with a period 
  $P_3 = 42.3\pm0.8$ years. An apsidal motion rate $\dot\omega = 0.0193\pm0.0010$ deg/cycle was obtained together with an 
  orbital eccentricity $e = 0.0548\pm0.0015$.  
  TESS data provide a value $T_2-T_1 = 1.4873\pm0.0005$ days and, adopting the eccentricity given by \cite{Clausen1996}, the 
  argument of periastron comes out to be $\omega = 204.9\pm2.2$ deg. When relating it with the result of the light curve analysis 
  in \cite{Clausen1996}, $\omega$ = 125.1$\pm$1.0 deg, an apsidal motion rate $\dot\omega = 0.0196\pm0.0006$ deg/cycle can be 
  derived, in agreement with all previous results.
  
  \item  DI Her
  
  This system has historically played a key role as a paradigmatic case for relativistic apsidal motion tests. Early apsidal motion 
  rate measurements showed a conspicuous and disturbing disagreement with theoretical predictions. This was resolved by 
  the work of \cite{Albrecht2009}, who found a large misalignment of the spin axes of the component stars from measuring 
  the Rossiter-McLaughlin effect. The most recent apsidal motion determination for DI Her, by \cite{Claret2010a}, yielded a 
  value $\dot\omega = 0.00042\pm0.00012$ deg/cycle, in good agreement with the predicted value when the observed rotational axes 
  misalignment is considered. Precise values of $T_2-T_1$ can be retrieved from the TESS measurements, as given in 
  Table\,\ref{tab:t2t1}. Using values from the literature, and restricting to photoelectric measurements only, the increase of the 
  observed separation in $T_2-T_1$ gives a slope of $(9.3\pm0.5) \times 10^{-6}$ days/cycle using a weighted least-squares minimization. 
  Assuming an orbital eccentricity $e = 0.489\pm0.003$, as given by \cite{Torres2010}, the observed apsidal motion rate of DI Her is 
  found to be $\dot\omega = 0.000435\pm0.000023$ deg/cycle, in good agreement but significantly more precise than previous studies.
  
  \item  EP Cru
  
  A first light curve of this eclipsing binary was obtained by \cite{Clausen2007}, while \cite{Albrecht2013} determined precise 
  physical parameters with the aim of studying circularization and synchronization timescales as well as a possible misalignment of the 
  rotational axes. With improved absolute dimensions, \cite{Albrecht2013} could predict the presence of apsidal motion but could 
  not measure it. They also pointed out the rotational velocities of the component stars, much larger than their synchronized values.
  We have used the precise value $T_2-T_1 = 6.8140\pm0.0002$ days given by \cite{Clausen2007} and compared it with the 
  values obtained from the TESS measurements. A weighted mean of the time differences in Table\,\ref{tab:t2t1} yields 
  $T_2-T_1 = 6.8255\pm0.0005$ some 1000 orbital cycles later. This is an increase rate of $(1.16\pm0.05) \times 10^{-5}$ days/cycle. 
  Assuming the orbital eccentricity given by \cite{Albrecht2013}, $e = 0.1874\pm0.0005$, we obtain $\dot\omega = 
  0.00237\pm0.00015$ deg/cycle, which becomes the first detection of apsidal motion in EP Cru.
  
  \item V760 Sco
  
  This system was studied by \cite{Andersen1985}, who obtained precise physical parameters and measured an apsidal motion 
  $\dot\omega = 0.0430\pm0.0017$ deg/cycle with an orbital eccentricity $e=0.0265\pm0.0010$. \cite{Wolf2000} re-discussed the 
  apsidal motion parameters with new eclipse timings and obtained $\dot\omega = 0.0443\pm0.0004$ deg/cycle, with $e = 0.0270\pm0.0005$. 
  Nevertheless, the new observations were not sufficiently precise for an improved determination of the eccentricity.
  From the TESS measurements, we obtain a value $T_2-T_1 = 0.8925\pm0.0005$ days, which is in good agreement 
  with the predicted value by the solution of \cite{Andersen1985} of $T_2-T_1 = 0.8919\pm0.0025$ days, but not 
  with the value predicted from \cite{Wolf2000} of $T_2-T_1 = 0.8946\pm0.0004$ days. 
  Considering all the possible $T_2-T_1$ values retrieved from \cite{Andersen1985}, computed with individual timings 
  within less than 10 orbital cycles, and adopting their value of the eccentricity, we obtain the corresponding arguments 
  of periastron at different epochs. We adopted for the argument of periastron corresponding to the TESS value of $T_2-T1$, a 
  negative value of $e\sin\omega$. This conclusion was achieved by fitting the light curve fixing the binary parameters to those 
  reported in \cite{Andersen1985}. A weighted linear least-squares fit yields an apsidal motion rate of 
  $\dot\omega = 0.0434\pm0.0005$ deg/cycle, which includes the uncertainty in the orbital eccentricity.
  
  \item  MU Cas
  
  A complete photometric and spectroscopic study of MU Cas was carried out by \cite{Lacy2004}. The system shows a well-defined 
  orbital eccentricity of $e=0.1930\pm0.0003$, and the phase of the secondary eclipse is observed at $0.61914\pm0.00015$. 
  The argument of periastron obtained from the light curve analysis is $\omega = 13.4\pm0.4$  deg, but the small time  span of eclipses 
  available did not allow the authors to claim the detection of apsidal motion. Moreover, the argument of periastron close 
  to 0 deg, which makes it difficult to observe changes in timing differences, and, also, apsidal motion is expected to be very slow 
  given the observed relative radii. The measurements of $T_2-T_1$ from TESS give a phase of the secondary eclipse of 
  $0.61881\pm0.00003$, yielding an poorly-significant determination of the apsidal motion rate $\dot\omega = 0.0010\pm0.0005$ 
  deg/cycle. To obtain this value we adopted the orbital eccentricity from \cite{Lacy2004}. The alternative method, that is, using 
  the measured times of eclipse in Table 1 of \cite{Lacy2005} to calculate $T_2-T_1$ differences, provides $\dot\omega = 0.0013\pm0.0003$ 
  deg/cycle. The alternative method of comparing the argument of periastron in table 4 of \cite{Lacy2004} and that derived from 
  TESS phase of the secondary eclipse ($\omega = 14.00\pm0.05$ deg) yields $\dot\omega = 0.0009\pm0.0006$ deg/cycle. The 
  low statistical significance of the measurement, although consistent using different techniques, does not allow the use of MU Cas to
  test internal stellar structure. We derive a log k$_2$ value with a large uncertainty of 0.14 when using the observed apsidal 
  motion rate and after subtracting the relativistic contribution. 
  
  \item  GG Lup
  
  This system was studied by \cite{Andersen1993}, who obtained precise absolute dimensions as well as a good determination 
  of the apsidal motion rate of $\dot\omega = 0.0181\pm0.0006$ deg/cycles and an orbital eccentricity $e=0.150\pm0.005$. 
  \cite{Wolf2005} revisited the apsidal motion determination and obtained results in good agreement with those of \cite{Andersen1993}, 
  with an orbital eccentricity $e=0.1545\pm0.0010$. The most recent study by \cite{Budding2015} confirmed the latter value of the 
  eccentricity with an apsidal motion rate $\dot\omega = 0.0172\pm0.0003$ deg/cycle. TESS observations show the expected 
  slow increase of the eclipse timing differences over the short time span covered, and yield an average value 
  $T_2-T_1=0.7561\pm0.0002$ days. This is not compatible with predictions using the eccentricity by \cite{Andersen1993}. 
  We have computed the argument of periastron corresponding to the $T_2-T_1$ values available from individual timings separated by
  less than 10 orbital cycles and adopting an orbital eccentricity $e=0.155\pm0.005$, as indicated by \cite{Wolf2005}. A linear least-squares 
  fit of the variation of $\omega$ with time yields $\dot\omega = 0.0173\pm0.0003$ deg/cycle, in excellent agreement 
  with the study by \cite{Budding2015}.
  
  \item $\zeta$ Phe
  
  This bright eclipsing binary system has been the subject of a recent study using TESS data by \cite{Southworth2020b}. The early 
  photometric analyses by \cite{Clausen1976} already showed a significant third light contribution and the spectroscopic 
  analysis by \cite{Andersen1983} allowed accurate absolute parameters to be determined. Apsidal motion was reported by 
  \cite{Gimenez1986} using the individual eclipse timings available and obtained $\dot\omega = 0.0373\pm0.0055$ deg/cycle with an 
  eccentricity $e=0.0113$ given by the photometric light curve analysis but not taking into account light-time effects. 
  \cite{Zasche2007} combined astrometric measurements with light-time modeling to study the orbit of the third body 
  and derived an apsidal motion rate of $\dot\omega = 0.028\pm0.001$ deg/cycle with an eccentricity of $e=0.0107\pm0.0020$ for 
  the close orbit as well as a period of $P_3=221$ years and an eccentricity of $e=0.366$ for the wide orbit of the outer third companion.
  The light curve analysis of \cite{Southworth2020b} provides precise absolute dimensions for $\zeta$ Phe, and confirms the presence of 
  third light. The orbital eccentricity was found to be $e=0.0116\pm0.0024$, in excellent agreement with previous studies. The argument 
  of periastron is $\omega=307\pm12$ deg. The comparison of this value with the light curve of \cite{Clausen1976}, almost 10\,000 cycles 
  before, indicates an apsidal motion rate of $\dot\omega = 0.031\pm0.003$ deg/cycle. Using the argument of periastron computed from 
  the $T_2-T_1$ values retrieved from \cite{Clausen1976}, and those from all the TESS observations, using $e = 0.0116\pm0.0024$, 
  we derive $\dot\omega = 0.0328\pm0.0006$ deg/cycle, which is in agreement with but more precise  than all the other results.
  
  \item  IQ Per
  
  This eccentric binary was studied by \cite{Lacy1985}, who provided absolute dimensions after the preliminary values by 
  \cite{Hall1970} and derived an apsidal motion rate of $\dot\omega = 0.012\pm0.003$ deg/cycle for an orbital eccentricity
  $e=0.075\pm0.006$, derived from the radial velocity curve. The apsidal motion rate was revised by \cite{Degirmenci1997} to  
  $\dot\omega = 0.0141\pm0.0008$ deg/cycle, using an eccentricity $e=0.076\pm0.004$, and by \cite{Wolf2006}, obtaining 
  $\dot\omega = 0.0138\pm0.0003$ deg/cycle, using $e = 0.0763\pm0.0008$.
  TESS measurements yield a separation between primary and secondary eclipses of $T_2-T_1 = 0.79827\pm0.00011$ days 
  that does not agree with the previously mentioned apsidal motion solutions. Using the most recent determination, by \cite{Wolf2006}, the 
  argument of periastron would be $\omega = 178\pm2$ deg at the TESS epoch and the predicted time separation is 
  $T_2-T_1 = 0.78755\pm0.00007$ days, a significant difference of $0.01072\pm0.00012$ days. To solve this, we performed a fit
  of the high-precision TESS light curve and obtained $e = 0.0662\pm0.0005$ and $\omega = 182\pm3$ deg, with the physical parameters being 
  in agreement with those found by \cite{Lacy1985}. Precise  
  measurements of the time of the secondary eclipse are difficult to obtain and only a few are available in the literature. Adopting the 
  eccentricity derived from the TESS light curve analysis, we calculated
  the argument of periastron corresponding to the observed values of $T_2-T_1$ in the literature, restricting to photoelectric data and
  with primaries and secondaries within 10 orbital cycles. A linear fit yielded an apsidal motion rate $\dot\omega = 0.0150\pm0.0005$ deg/cycle, 
  which we adopted for Table\,\ref{tab:apsidal}.
  
  \item  PV Cas
  
  Precise absolute dimensions of PV Cas are based on the studies by \cite{Popper1987} and \cite{Barembaum1995}. An apsidal motion 
  $\dot\omega = 0.0189\pm0.0015$ deg/cycle was measured by \cite{Gimenez1982} assuming an eccentricity value $e=0.0322\pm0.0005$. The most 
  recent subsequent study by \cite{Svariek2008} gives an apsidal motion rate $\dot\omega = 0.01896\pm0.00014$ deg/cycle with an orbital 
  eccentricity $e=0.03248\pm0.00014$. The TESS timing data alone indicate the presence of apsidal motion by comparing the measurements 
  of $T_2-T_1$ in the two available sectors, separated by 110 orbital cycles. Adopting an eccentricity $e=0.032$, the observed 
  variation yields $\dot\omega = 0.019\pm0.005$ deg/cycle but with poor precision. The more precise apsidal motion solution by 
  \cite{Svariek2008} predicts an argument of periastron of $\omega = 21.7\pm0.6$ deg at the time of the TESS observations, 
  equivalent to an eclipse separation $T_2-T_1 = 0.9090\pm0.0004$ days, which differs from the TESS measurements by $0.0036\pm0.0005$ days. 
  This disagreement prompted us to reanalyze the apsidal motion determination on the basis of the observed $T_2-T_1$ values using all 
  photoelectric data available in the literature, with individual timings separated by no more than 10 orbital cycles. Adopting an orbital 
  eccentricity of $e=0.0325\pm0.0005$ as given by \cite{Svariek2008} but with a larger, more realistic, uncertainty, the corresponding argument 
  of periastron is calculated for each value of $T_2-T_1$. The value of $\omega$ for TESS, close to 360 deg, was checked with an analysis of 
  the full light curve. A linear fit yielded $\dot\omega = 0.0212\pm0.0002$ deg/cycle, which we adopt 
  for Table\,\ref{tab:apsidal}. Concerning the argument of periastron at the time of the 
  observations by \cite{Ibanoglu1974}, we did not include the value of $\omega$ = 180$\pm$8 deg from their light curve analysis due to the uncertain determination of $e\sin\omega$. Nevertheless, our solution predicts a value in excellent agreement of $\omega = 175.8\pm1.64$ deg.
  
  \item  V364 Lac
  
  A detailed spectroscopic and photometric study of this A-type eclipsing binary was performed by \cite{Torres1999}. The authors 
  could detect the presence of apsidal motion in V364 Lac at a rate of $\dot\omega = 0.00258\pm0.00033$ deg/cycle from the combined 
  study of the radial velocity curve and eclipse timings, with an orbital eccentricity $e=0.2873\pm0.0014$. More recently, 
  \cite{Bulut2013} adopted the same value of the eccentricity and obtained $\dot\omega = 0.00207\pm0.00034$ deg/cycle. 
  The TESS data indicate a position of the secondary eclipse with respect to the primary of $T_2-T_1 = 3.73870\pm0.00005$ days, 
  which yields an argument of periastron $\omega = 87.43\pm0.02$ deg when adopting the eccentricity of \cite{Torres1999}. Comparing with 
  the value in table 3 of \cite{Torres1999}, an apsidal motion rate $\dot\omega = 0.00181\pm0.00006$ deg/cycle is derived, where the 
  uncertainty in the eccentricity is considered.
  
  \item  SW CMa
  
  \cite{Lacy1997} first determined the absolute dimensions of this relatively evolved DLEB and indicated the possible 
  presence of apsidal motion. \cite{Clausen2008} obtained a better-covered light curve and measured a rate 
  $\dot\omega = 0.00067\pm0.00021$ deg/cycle. Finally, \cite{Torres2012} reanalyzed the light curve 
  of \cite{Clausen2008}, together with newly obtained radial velocities, and obtained precise physical parameters.
  SW CMa was observed by TESS in two sectors separated by 60 orbital cycles. The variation in $T_2-T_1$ over this time interval is 
  too small to allow determining a apsidal motion rate value. The average timing difference is $T_2-T_1 = 3.1141\pm0.0002$ days. 
  The apsidal motion parameters of \cite{Clausen2008} predict an argument of periastron $\omega = 164.32\pm0.30$ deg while, 
  adopting the eccentricity from the same authors with an uncertainty $\pm$0.0005, the TESS $T_2-T_1$ yields $\omega = 164.34\pm0.30$ deg. 
  The excellent agreement shows that the apsidal motion rate given by \cite{Clausen2008} is accurate. However, the uncertainty could 
  be reduced thanks to the wider time span. A comparison of the argument of periastron derived from the TESS measurements with 
  the light curve analyses by \cite{Lacy1997} and \cite{Clausen2008} does not allow to improve the determination due to the very 
  slow apsidal motion rate and the proximity of $\omega$ to 180 deg. We analyzed the light curve obtained by TESS and obtained 
  a best-fitting $e = 0.3180\pm0.0005$. A fit using the observed values of $T_2-T_1$ with TESS, together with those 
  provided by \cite{Clausen2008}, yields an apsidal motion rate $\dot\omega = 0.00069\pm0.00005$ deg/cycle considering 
  all the uncertainties involved, which improves on the precision of the previous determination.
  
  \item  PT Vel
  
  \cite{Bakis2008} studied this system and derived the physical properties of their components as well as 
  the apsidal motion rate. Together with the orbital eccentricity of $e=0.127\pm0.006$, the authors obtained $\dot\omega = 0.0099\pm0.0003$ 
  deg/cycle. Nevertheless, these results are based on the light curves obtained by the ground-based All-Sky Automated Survey (ASAS) over a 
  decade, with poor coverage of the secondary eclipse. We have analyzed the TESS light curve and obtained a much better fit with a 
  lower eccentricity $e = 0.112\pm0.003$, but with the same relative radii and orbital inclination, thus with no change in the physical 
  parameters. Adopting the eccentricity derived from the TESS light curve, the $T_2-T_1$ measurements already show the presence of 
  apsidal motion. To increase the time span, we have compared the average eclipse time difference of the TESS measurements, 
  $T_2-T_1 = 1.00687\pm0.00012$ days, with the values obtained from the ASAS survey given by \cite{Kim2018}. A linear fit to the corresponding values
  of $\omega$ yields an apsidal motion rate of $\dot\omega = 0.0125\pm0.0006$ deg/cycle. The significant difference of the eccentricity, 
  determined from the TESS light curve or the ASAS analysis, that we could not reproduce, remains unexplained.
  
  \item  V1647 Sgr
  
  The absolute dimensions of V1647 Sgr were obtained by \cite{Andersen1985}, who could also measure an apsidal motion rate of 
  $\dot\omega = 0.00546\pm0.00006$ deg/cycle with an orbital eccentricity $e=0.4130\pm0.0005$. The authors also confirmed that the 
  visual companion of V1647 Sgr is physically bound. \cite{Wolf2000} re-analyzed the apsidal motion and obtained a rate 
  $\dot\omega = 0.00609\pm0.00012$ deg/cycle with an eccentricity $e=0.4142\pm0.0011$. The TESS data provide an average position 
  of the secondary eclipse with respect to the primary of $T_2-T_1 = 1.0880\pm0.0003$ days. The argument of periastron calculated for 
  the TESS measurements and those obtained from the individual timings in table 4 of \cite{Andersen1985} show a well-defined 
  variation of $\omega$. A linear fit yields an apsidal motion rate of $\dot\omega = 0.00554\pm0.00005$ deg/cycle, 
  considering the value and uncertainty in $e$ from \cite{Andersen1985}.
  
  \item  AI Hya
 
  First absolute dimensions for this system were published by \cite{Popper1998} on the basis of  their  own radial velocity curve 
  and the well-covered light curve by \cite{Joergensen1978}. An orbital eccentricity $e=0.230\pm0.002$ was 
  derived from the light curve, and the existence of pulsations in the secondary component was reported but no apsidal 
  motion. AI Hya has recently been studied by \cite{Lee2020}, who analyzed the light curve observed by TESS to characterize 
  the $\delta$~Scuti-type pulsations. The authors also analyzed the displacement of the argument of periastron and 
  derived a low-significance apsidal motion rate using individual times of minimum of $\dot\omega = 0.0017\pm0.0007$ deg/cycle 
  with an orbital eccentricity $e=0.241\pm0.083$. The more precise value derived from their own light curve analysis, 
  $e=0.234\pm0.002$, was not used. We have computed the $T_2-T_1$ values corresponding to individual eclipse timings within less than 
  10 orbital cycles from table 1 of \cite{Lee2020}, including the TESS measurements in our Table 2, and determined the argument of periastron 
  for each of them, adopting $e = 0.234\pm0.002$. A linear fit yields $\dot\omega = 0.00191\pm0.00005$ deg/cycle.
  
  \item  VV Pyx
  
  This system has kept the name of VV Pyx in spite of its re-naming as V596 Pup in the 78$^{\rm th}$ name-list of variable 
  stars \citep{Kazarovets2006}. A detailed photometric and spectroscopic study of the system by \cite{Andersen1984} 
  yielded precise absolute dimensions and an apsidal motion rate $\dot\omega = 0.00142\pm0.00045$ deg/cycle was obtained with the orbital 
  eccentricity derived from the light curve analysis, $e = 0.0956\pm0.0009$. A significant contribution from third light was measured 
  and attributed to the close visual companion of the eclipsing binary but no effect in the measured apsidal motion was measured. 
  TESS data alone indicate a small variation of the $T_2-T_1$ values with time, corresponding to an apsidal motion rate of 
  $\dot\omega = 0.00138\pm0.00012$ deg/cycle. Using also the two additional $T_2-T_1$ values derived from table 5 in \cite{Andersen1984}, 
  we obtained a final rate $\dot\omega = 0.00132\pm0.00005$ deg/cycle, using $e = 0.0956\pm0.0009$.
  
  \item  EK Cep
  
  \cite{Popper1987} was the first to identify the pre-main sequence nature of the secondary component in EK Cep. Radial velocities 
  were measured by \cite{Tomkin1983} and the light curve was analyzed by \cite{Hill1984}. An orbital eccentricity 
  $e=0.109\pm0.003$ is derived from the radial velocity curve and adopted for the photometric studies. \cite{Gimenez1985b} 
  obtained an apsidal motion rate $\dot\omega = 0.00107\pm0.00032$ deg/cycle using eclipse timings. TESS measurements provide a 
  value $T_2-T_1 = 2.3988\pm0.0002$ days, and we have used all available values of $T_2-T_1$ from the literature with precision 
  better than 0.001 days, which are challenging because of the shallow secondary eclipse. With these values, we obtained a slope of 
  $(-3.61\pm0.08) \times 10^{-6}$ days/cycle, from which an apsidal motion rate of $\dot\omega = 0.00088\pm0.00004$ deg/cycle is derived, 
  adopting $e = 0.109\pm0.003$. 
  
  \item  VV Crv
  
  \cite{Fekel2013} studied this bright system combining spectroscopic and photometric observations, and obtained the 
  physical parameters given in Table 1. Conspicuous third light was observed, which complicated the light curve analysis. Moreover, 
  the long duration and shallow depth of the eclipses make their timing difficult, and no apsidal motion had been reported. From the 
  TESS measurements we obtain an average eclipse difference $T_2-T_1 = 1.559\pm0.004$ days. The corresponding argument 
  of periastron, adopting an eccentricity $e=0.0852\pm0.0010$ given by \cite{Fekel2013}, is $\omega=265.5\pm1.5$ deg with its 
  uncertainty dominated by that of the eclipse timings. Given the observed dispersion in the values of $T_2-T_1$, we performed an 
  analysis of the TESS light curve with the same orbital eccentricity, and obtained an argument of periastron $\omega = 266.5\pm0.5$ deg, 
  in agreement with the previous value but more precise. The argument of periastron given by \cite{Fekel2013} in their table 7, 
  from the light and velocity curve combined analysis, is $\omega = 257.7\pm0.2$ deg, thus indicating for the first time the presence of 
  apsidal motion in VV Crv. The change in omega, from the light curve fits, after 807 orbital cycles, gives an apsidal motion 
  rate $\dot\omega = 0.0109\pm0.0012$ deg/cycle. This result should be nevertheless considered preliminary due to the observed dispersion 
  in the eclipse timings, and difficulties with the photometric analysis, due to the presence of third light. 
  
  \item  IM Per
  
  Absolute properties of IM Per were obtained by \cite{Lacy2015} and revised recently by \cite{Lee2020} using the TESS light curve. 
  A simultaneous radial velocity and timing analysis carried out by \cite{Lacy2015} gave an apsidal motion rate of 
  $\dot\omega = 0.0147\pm0.0008$ deg/cycle, while a value $\dot\omega = 0.0155\pm0.0008$ deg/cycle was obtained by the same authors from 
  fitting the light curve with variable $\omega$. In both cases, the orbital eccentricity was fixed to $e=0.047\pm0.003$. The TESS 
  light curve solution by \cite{Lee2020} gives a compatible but slightly larger value at $e = 0.0491\pm0.0010$. Using all the $T_2-T_1$ 
  measurements available from table 1 of \cite{Lacy2015} plus the value from TESS $T_2-T_1 = 1.09484\pm0.00014$ days, 
  we calculated the corresponding argument of periastron adopting this eccentricity. A linear fit to the variation 
  of $\omega$ with time then yields $\dot\omega = 0.0146\pm0.0004$ deg/cycle. Considering only the timings used by 
  \cite{Lacy2015}, a rate of 0.0149$\pm$0.0005 deg/cycle is obtained. 
  
  \item  BP Vul
  
  \cite{Lacy2003} obtained precise absolute dimensions for BP Vul, an eccentricity $e=0.0355\pm0.0027$, and the 
  secondary eclipse at phase $0.47968\pm0.00021$. The authors also analyzed the apsidal motion rate but was found to be 
  negative and the potential presence of a perturbing third body was argued to explain this anomaly, but the timespan 
  covered was too narrow for a conclusion. \cite{Csizmadia2009} revised the situation with the analysis of an earlier, 
  unpublished, light curve showing a positive apsidal motion with a rate of 1 deg/year but with only one new secondary 
  eclipse timing. Moreover, their analysis of individual timings requires a light-time effect that is not 
  quantified. On the other hand, TESS measurements give a phase for the secondary eclipse of 
  $0.47989\pm0.00010$ days. Given that this value is not significantly different from the light curve by \cite{Lacy2003} no claim of apsidal 
  motion detection can be made at this time. 
  
  \item  V1022 Cas
  
  A recent analysis of this eclipsing system by \cite{Southworth2021}, using TESS data, provides precise physical parameters 
  in good agreement with those published by \cite{Lester2019} on the basis of a combined astrometric and spectroscopic 
  study. The lack of precise eclipse timings and the small time separation with respect to the TESS measurements makes 
  the detection of apsidal motion very difficult. We have used the spectroscopic solution of \cite{Lester2019}, $e$ and $\omega$, 
  in their table 4 to obtain the position of the secondary eclipse at phase $0.66565\pm0.00009$. Comparing this value 
  with the TESS measurements from the light curve solution by \cite{Southworth2021}, 452 orbital cycles later, gives a phase 
  displacement of $-0.00026\pm0.00011$, equivalent to an apsidal motion rate $\dot\omega = 0.00032\pm0.00015$ deg/cycle. 
  When subtracting the relativistic term, the classical term comes out to be $\dot\omega = 0.00007\pm0.00015$ deg/cycle, which 
  cannot be used in the present paper. 
  
  \item  PV Pup
  
  Despite the presence of intrinsic variability of unknown origin, the light curve obtained by \cite{Vaz1984} provides 
  well-defined photometric elements, and their combination with radial velocity measurements yield precise physical 
  properties of the component stars. The orbital eccentricity was found to be $e=0.0503\pm0.0011$ but no apsidal motion was reported. 
  PV Pup belongs to the stellar system ADS 6348 but is most probably an optical pair. TESS measurements of the separation 
  between primary and secondary eclipses present a large dispersion, higher than the individual errors, probably due to the 
  observed intrinsic variability. The average values of $T_2-T_1$ for two sectors, separated by 445 orbital cycles, allowed 
  us to calculate the corresponding argument of periastron with the eccentricity given by \cite{Vaz1984}, and thus 
  estimate an apsidal motion rate of $\dot\omega = 0.0087\pm0.0010$ deg/cycle. Given the noise in the eclipse timings, also noticed by 
  \cite{Vaz1984}, we tried to analyze the light curve of TESS with $e = 0.0503$. A significant fraction of third light 
  was observed and the intrinsic variability did not allow to achieve a reasonable fit. Leaving only $\omega$ as a free parameter, 
  the best fit indicated, when compared with the value of the light curve analysis by \cite{Vaz1984}, an apsidal motion 
  rate of $\dot\omega = 0.0072\pm0.0005$ deg/cycle. Therefore, the preliminary result with only the TESS timings cannot be 
  confirmed and we call for further monitoring of PV Pup before establishing its apsidal motion rate.
  
  \item  BF Dra
  
  Apsidal motion was determined by \cite{Wolf2010} to be $\dot\omega = 0.00056\pm0.00008$ deg/cycle but leaving free the eccentricity to converge 
  to $e = 0.3898$ despite the poor coverage of the apsidal motion variations. Masses and radii for the components of BF Dra were obtained 
  by \cite{Lacy2012b}, who also analyzed the apsidal motion by fitting the individual eclipse timings and obtained 
  $\dot\omega = 0.00049\pm0.00008$ deg/cycle with an eccentricity $e=0.3865\pm0.0005$. BF Dra was observed in 12 TESS sectors, spanning a 
  total of 352 days, and the large number of precise values of $T_2-T_1$ allowed us to determine its apsidal motion rate independently of any 
  previous measurement. A linear fit to the TESS values of $T_2-T_1$ in Table\,\ref{tab:t2t1} yields a slope of the variation of
  $(-2.19\pm0.14) \times 10^{-6}$ days/cycle. Adopting the eccentricity obtained by \cite{Lacy2012b}, this gives an apsidal motion of 
  $\dot\omega = 0.00042\pm0.00003$ deg/cycle. Because of the difference with previous results, we calculated separately the linear periods 
  for primary and secondary eclipses and obtained a difference $\Delta P = (-2.26\pm0.04) \times 10^{-5}$ days/cycle, but with systematics 
  present in the residuals of both primary and secondary TESS timings. We interpret these as potentially due to the light travel-time 
  effect induced by a third body orbiting the system. \cite{Lacy2012b} did indeed identify the presence of third light in the light curve of 
  BF Dra. The effect of a third body, however, should not hamper the apsidal motion determination using the $T_2-T_1$ values. From the 
  archival times of minimum we could retrieve five precise $T_2-T_1$ measurements, considering only individual eclipse timings within 10 
  orbital cycles. When adding these to the TESS $T_2-T_1$ measurements the slope becomes $(-2.35\pm0.05) \times 10^{-5}$ days/cycle 
  and $\dot\omega = 0.00045\pm0.00002$ deg/cycle, compatible with the values resulting from the TESS measurements alone and from 
  \cite{Lacy2012b}. We detected some remaining systematic residuals in the TESS observations that remain unexplained. An alternative 
  determination method is the use of the value of $\omega$ from the light curve analysis, rather than the times of eclipse. Comparing the argument 
  of periastron given by \cite{Lacy2012b} in their table 7 with that computed from the average TESS $T_2-T_1$ measurements, 
  430 orbital cycles later, a small but significant increase in $\omega$ of $0.16\pm0.03$ deg is observed. This yields an apsidal motion rate of 
  $\dot\omega = 0.00037\pm0.00007$ deg/cycle, which is in marginal agreement with the value described before. More observations are needed 
  to obtain a definitive apsidal motion rate as well as analyzing the impact of the third body. In any case, the large fractional relativistic 
  term (68\%) of BF Dra implies a classical term with an uncertainty of $\sim$30\% thus rendering this system unsuitable for the comparison 
  with theoretical stellar models.
  
  \item V1143 Cyg
 
  \cite{Andersen1987} obtained absolute dimensions for V1143 Cyg, which were more recently revised by \cite{Lester2019} with 
  new observations. Apsidal motion has been studied by \cite{Khaliullin1983}, \cite{Gimenez1985b}, \cite{Burns1996}, \cite{Dariush2005} and, 
  finally, by \cite{Wolf2010}. The latter authors gave $\dot\omega = 0.00072\pm0.00008$ deg/cycle with an eccentricity $e=0.535\pm0.004$. 
  A fit to the TESS timing data alone yields a faster rate at $\dot\omega = 0.00088\pm0.00005$ deg/cycle. A wider time span can
  be obtained by using the individual timings from \cite{Wolf2010}, with their original errors when available or otherwise adopting an
  uncertainty of $\pm$0.005 days, that we judged realistic. The residuals of 
  the fit to all timing differences show a significantly larger dispersion than the uncertainties of the archival 
  measurements, suggesting an underestimation of the errors. However, we assumed that they are globally unbiased and 
  obtained a slope of the $T_2-T_1$ variation of $(2.4\pm0.1) \times 10^{-5}$ days/cycle. Adopting the orbital 
  eccentricity given by \cite{Lester2019}, $e = 0.5386\pm0.0004$, we obtain an apsidal motion for V1143 Cyg of 
  $\dot\omega = 0.00080\pm0.00004$ deg/cycle. 
  
  \item  IT Cas
  
  Precise physical properties of the components of IT Cas were obtained by \cite{Lacy1997}. From their light and radial velocity curves 
  the authors obtained an orbital eccentricity $e=0.085\pm0.004$ and a longitude of periastron of $\omega = 332.6\pm1.4$ deg, but 
  apsidal motion could not be determined due to the short time span covered with photoelectric measurements. Moreover, the same authors 
  obtained different values of the eccentricity from the radial velocity, the light curve and the ephemeris solution. 
  \cite{Kozyreva2001} used their 
  own measurements together with those by \cite{Lacy1997} to obtain $e = 0.089\pm0.002$ with an apsidal motion rate
  $\dot\omega = 0.0012\pm0.0003$ deg/cycle, comparing the values of $\omega$ from the different light curve solutions. TESS 
  timings confirm the values of \cite{Kozyreva2001}. We have combined the TESS 
  eclipse timing differences with those derived from the most precise individual timings listed in table 1 of \cite{Kozyreva2001}, only 
  photoelectric and within 10 orbital cycles. The fit provides a slope of $(2.01\pm0.05)\times10^{-6}$ days/cycle. Adopting $e = 0.089\pm0.002$, 
  yields an apsidal motion rate $\dot\omega = 0.00114\pm0.00010$ deg/cycle.
  
  \item  AI Phe
  
  The light curve obtained by \cite{Hrivnak1984} was analyzed by \cite{Andersen1988} together with their own 
  radial velocity curve. \cite{Maxted2020} analyzed the light curve provided by TESS and improved the physical parameters 
  using the radial velocity measurements by \cite{Helminiak2009}. TESS measurements indicate a precise position of the 
  secondary eclipse at phase $0.457865\pm0.000005$ and an argument of periastron $\omega = 110.34\pm0.11$ deg. Unfortunately, 
  accurate eclipse timings are difficult to obtain due to the long orbital period and the duration of the eclipses. 
  \cite{Kirkby2016} showed significant variations in the sidereal period, probably due to the presence of a third body, and 
  determined the phase of the secondary eclipse at $0.4584\pm0.0015$, which does not differ significantly from the TESS measurement. 
  On the other hand, the light curve solution by \cite{Andersen1988} gives a value of $\omega = 109.6\pm1.0$ deg, which is again
  very similar to the value from the TESS light curve. More eclipse timings are certainly needed, as well as a complete 
  dynamical study to evaluate the impact of the potential third body in the measurement of apsidal motion.
  
  \item EW Ori
  
  This DLEB was analyzed by \cite{Clausen2010}, who obtained the physical parameters of its components
  and estimated an apsidal motion rate $\dot\omega = 0.00042\pm0.00010$ deg/cycle with an orbital eccentricity $e=0.0758\pm0.0020$ 
  derived from the light curve. We obtained precise $T_2-T_1$ values from the TESS light curve. In addition, we retrieved four 
  $T_2-T_1$ values from the literature \citep{Wolf1997,Clausen2010} from the evaluation of the secondary phase displacement and  
  using only precise individual timings separated by less than 10 orbital cycles. A linear fit to the $T_2-T_1$ 
  variation with time yields a slope of $(1.48\pm0.11)\times10^{-6}$ days/cycle. Taking the orbital eccentricity given by 
  \cite{Clausen2010} produces an apsidal motion rate of $\dot\omega=0.00033\pm0.00002$ deg/cycle. Unfortunately, this system cannot be used
  to compare with stellar structure models. EW Ori´s  highly relativistic periastron precession  is dominant (79\%) and the determination 
  of the classical term carries an uncertainty that is above our threshold. However, this system could be added to our sample in
  paper I. Following the methodology there, we determine $\dot\omega_{\rm GR} = 0.00026\pm0.00002$ deg/cycle after subtracting the 
  classical term calculated from the models, as described in section 4. It can be shown that the measured relativistic rate is in very 
  good agreement with the other systems plotted in figures 10 and 11 of paper I. A recalculation of the corresponding post-Newtonian 
  parameters, now including EW Ori, yields $A = 1.000\pm0.011$ and $B = 0.000\pm0.051$. 
 
  \item  V530 Ori
  
  A complete analysis of this G+M system was carried out by \cite{Torres2014}, who provided the physical parameters of its components. 
  The shallow total secondary eclipse makes it difficult to study variations in phase. \cite{Torres2014} determined the orbital 
  eccentricity to be $e=0.0862\pm0.0010$ with $\omega = 130.08\pm0.14$ deg. Apsidal motion was detected by the authors but could not 
  be determined with sufficient significance. From their table 1, only two precise $T_2-T_1$ values could be retrieved with individual 
  timings within 10 orbital cycles. Using those with the new TESS data, we determine an apsidal motion rate 
  $\dot\omega = 0.00086\pm0.00005$ deg/cycle, with the orbital eccentricity of $e=0.0862\pm0.0010$. Nevertheless, additional observations
  are needed to confirm the apsidal motion rate, given the short span of time available and the difficulty to obtain precise timings.
  
 \end{itemize}

\end{appendix}

\end{document}